\documentclass[reprint,aps,amsmath,amssymb,superscriptaddress,twocolumn,longbibliography,prx]{revtex4-2}
\pdfoutput=1
\usepackage{graphicx,enumerate,mathtools,bm}
\usepackage[caption=false]{subfig}
\usepackage{floatrow}
\usepackage{xr-hyper}
\usepackage{hyperref}
\usepackage{natbib}

\hypersetup{
     colorlinks = true,
     citecolor = blue,
     urlcolor = blue,
     linkcolor = red
}

\makeatletter
\newcommand*{\addFileDependency}[1]{
	\typeout{(#1)}
	\@addtofilelist{#1}
	\IfFileExists{#1}{}{\typeout{No file #1.}}
}
\makeatother

\newcommand*{\myexternaldocument}[1]{%
	\externaldocument{#1}%
	\addFileDependency{#1.tex}%
	\addFileDependency{#1.aux}%
}

\myexternaldocument{LV_cavity_supplemental}

\begin{document}
	
	\title{Understanding Species Abundance Distributions in Complex Ecosystems of Interacting Species}
	\author{Jim Wu}
	\affiliation{Physics Department, Princeton University, Princeton, NJ}
	\affiliation{CUNY-Princeton Center for Biological Function}
	
    \author{Pankaj Mehta}
	\affiliation{Physics Department, Boston University, Boston, Massachusetts}

	\author{David Schwab}
	\affiliation{CUNY-Princeton Center for Biological Function}
	\affiliation{Initiative for Theoretical Sciences, The Graduate Center, City University of New York, New York, New York}
	
	\date{\today}
	
	\begin{abstract}
		
Niche and neutral theory are two prevailing, yet much debated, ideas in ecology proposed to explain the patterns of biodiversity. Whereas niche theory emphasizes selective differences between species and interspecific interactions in shaping the community, neutral theory supposes functional equivalence between species and points to stochasticity as the primary driver of ecological dynamics. In this work, we draw a bridge between these two opposing theories. Starting from a Lotka-Volterra (LV) model with demographic noise and random symmetric interactions, we analytically derive the stationary population statistics and species abundance distribution (SAD). Using these results, we demonstrate that the model can exhibit three classes of SADs commonly found in niche and neutral theories and found conditions that allow an ecosystem to transition between these various regimes. Thus, we reconcile how neutral-like statistics may arise from a diverse community with niche differentiation. 
    \end{abstract}
    \maketitle

\section{Introduction}
    The enduring challenge of community ecology is to understand the salient processes that shape the patterns of biodiversity observed across many ecosystems. From the work of MacArthur, ecological niches became a popular explanation of community structure and dynamics. In niche theory, each species has a unique set of traits that specializes them to particular resources and habitats. Through competition with other species, the ecosystem is partitioned into distinctive niches occupied by one species, thus allowing many species with differing traits to coexist within the same ecosystem. Hence, niche theory proposes that diversity arises from environmental heterogeneity and species interactions, highlighting the central role of selection in shaping ecosystems \cite{azaele_statistical_2016,fisher_transition_2014,haegeman_mathematical_2011,gravel_reconciling_2006,chave_neutral_2004,kalyuzhny_niche_2014,matthews_neutral_2014,grilli_laws_2019}. However, in recent decades, Hubbell's studies of island biogeography and biodiversity sparked an opposing neutral theory of ecology. Ecological neutrality starts from an assumption of functionally equivalent species, where species from the same trophic level compete for similar resources and share similar phenotypic traits \cite{azaele_statistical_2016,haegeman_mathematical_2011,gravel_reconciling_2006,chave_neutral_2004,matthews_neutral_2014,hubbell_unified_2001,matthews_neutral_2014}. To explain the coexistence of many species, neutral theory identifies ecological processes such as population drift, migration, and dispersal. These mechanisms are independent of species traits and capture the inherent randomness of ecological events and population dynamics. So, instead of niche differences and selection, neutral theory promotes stochasticity as the main driver of community assembly. 
    
\begin{figure}[t]
    \centering
    \includegraphics[width = 0.9\textwidth]{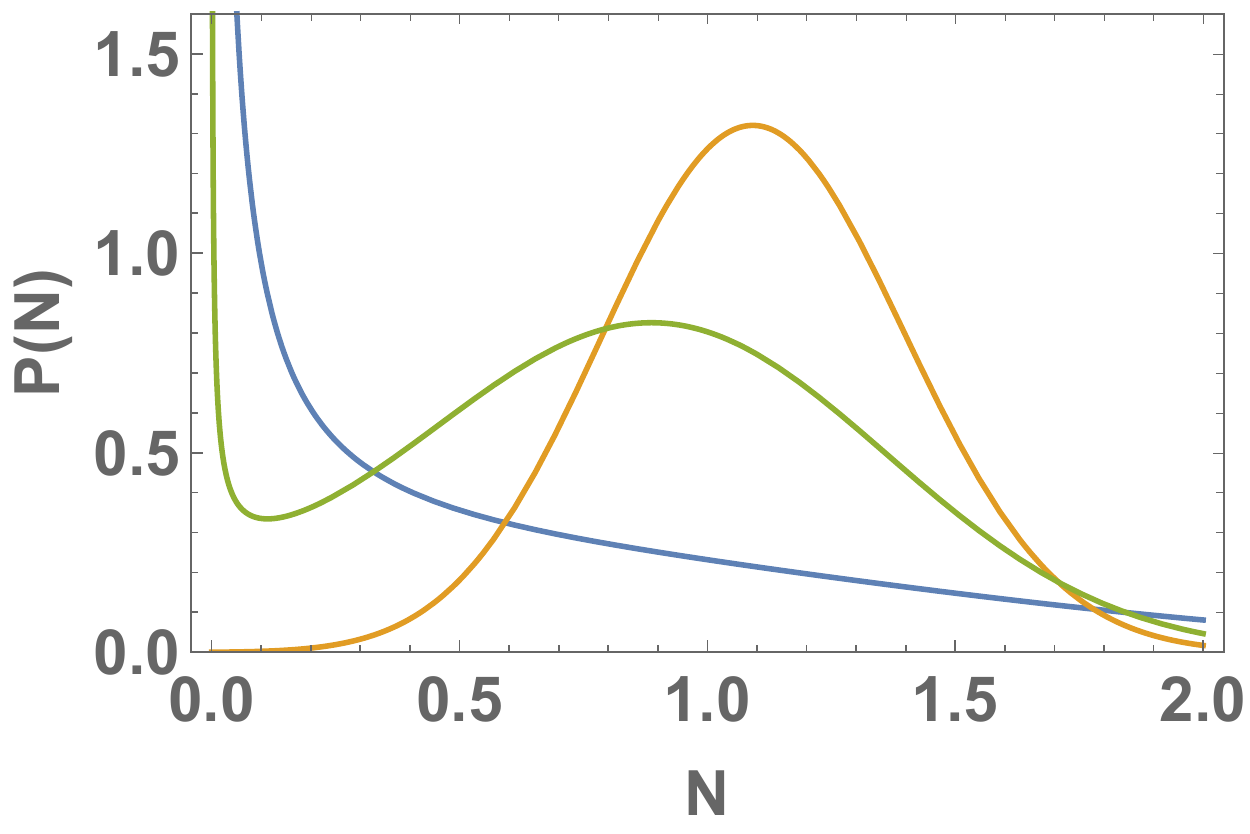}
    \caption{Common marginal distributions $\mathbf{P(N)}$, or species abundance distributions (SADs) found in real ecosystems and theoretical models. When the rate of immigration into the community is high compared to demographic noise strength, the SAD resembles a log-normal distribution (orange). However, if demographic noise dominates over migration, then there are two scenarios. A community dominated by noise has very few species surviving at high abundances and thus has an SAD that behaves like the Fisher log-series (blue). If selection is the primary driver of ecological dynamics, then a local maximum manifests at intermediate abundances (green). }
    \label{fig:SAD}
\end{figure}

   Both niche and neutral theory have found success in explaining ecological data. Although the latter has been criticized for its unrealistic assumption of functional equivalence and neglect of phenotypic differences observed between species, the minimal neutral model still reproduces many of the observed macro-ecological patterns and statistics \cite{azaele_statistical_2016,matthews_neutral_2014,chisholm_niche_2010}. One of the most common and fundamental patterns is the species abundance distribution (SAD), which quantifies the community structure by counting the number of species of a particular population size. Not only are SADs relatively easy to compute from data, they have been used to verify the ecological mechanisms that give rise to the observed structure of the community. Over the years, studies of real ecosystems and theoretical models have yielded numerous species abundance distributions, with the most popular being the Fisher log-series and the lognormal-like distributions \cite{azaele_statistical_2016,matthews_neutral_2014,chisholm_niche_2010}. Theoretically, species-rich metacommunities shaped by neutral processes are typically described by log-series distributions, which are characterized by many rare species and relatively few abundant species. On the other hand, lognormal-like SADS (such as the zero-sum multinomial distributions) are usually found in local communities with migration from a species pool \cite{azaele_statistical_2016,matthews_neutral_2014}. Instead of a decrease in number of species as abundance increases, these lognormal-like SADs have very few species at extremely low and high abundances and a distinct peak at intermediate population sizes. Plots of these common classes of neutral SADs are shown in Figure \ref{fig:SAD}.
   
   It is accepted nowadays that niche and neutral theory are not dichotomous paradigms of ecology, but are rather frameworks that only emphasize the importance of one set of processes in shaping community assembly. In real ecosystems, both niche and neutral mechanisms are vital and must be incorporated into a unified framework. \cite{fisher_transition_2014,haegeman_mathematical_2011,gravel_reconciling_2006,chave_neutral_2004}. To reconcile niche and neutral theories, ecologists have been investigating how neutral statistics might emerge from a community model with niche differences without the assumption of functional equivalence. In these models integrating both niche and neutral processes, one can tune the model parameters to continuously traverse between regimes dominated by selective differences and those dominated by stochasticity \cite{azaele_statistical_2016,haegeman_mathematical_2011,gravel_reconciling_2006,noble_multivariate_2011}. More recently, some ecologists have proposed the idea of emergent neutrality (EN) \cite{azaele_statistical_2016,matthews_neutral_2014,scheffer_self-organized_2006,scheffer_toward_2018,barabas_emergent_2013}. Starting from an ecosystem of ecologically distinctive species, interspecific interactions and evolution can drive the community to exhibit clusters of ecologically similar species. Although only one species in each cluster ultimately survives, similar species can transiently coexist for many generations. Hence, EN unifies niche and neutral theory by proposing neutrality as an emergent property of ecosystems shaped by niche differentiation. In addition to bridging the two theories, emergent neutrality has gained some traction as it produces SADs with multiple modes, which have been found across various communities ranging from phytoplankton to birds \cite{scheffer_self-organized_2006,scheffer_toward_2018,vergnon_emergent_2012,vergnon_interpretation_2013,barabas_emergent_2013}. An example of an SAD with a secondary maximum is shown in Figure \ref{fig:SAD}.

    From exploring the species diversity of numerous ecosystems, ecologists have found a collection of different species abundance distributions and have applied neutral models to explain the various forms of SADs. However, it still remains an issue to reconcile traditional niche theory with neutral theory. Here, we construct a stochastic Lotka-Volterra (LV) niche model of a biodiverse ecosystem that incorporates both random interactions between species and neutral processes such as demographic noise and migration. Within this stochastic LV model, we capture three main classes of species abundance distributions: log-series-like, log-normal-like, and bimodal. The boundaries between the different SAD behavior hinge on the balance of niche and neutral processes and the relative values of the macro-ecological parameters shaping ecological dynamics. Furthermore, we find that at high levels of noise, the neutral phase can occupy a large volume of the ecological ''phase'' diagram. Although setting the stochasticity parameter to high levels may be deemed unrealistic, we argue heuristically that this may actually correspond to typical realizations of real ecosystems. Not only does this demonstrate how neutral-like statistics can manifest in the presence of selective forces, but it also explains the prevalence of these neutral macro-ecological patterns observed across many ecosystems.

   \floatsetup[figure]{style=plain,subcapbesideposition=top}

\section{The Model}

	Consider a well-mixed community with a pool of $S$ different species, each with an abundance $\tilde{N}_i\geq 0$ where $i = 1,2,\ldots S$. The population size of each species is shaped by processes such as birth, death, migration, and interactions with other species. An effective model that captures the population dynamics of such ecosystem is the stochastic Lotka-Volterra equation, which reads
	\begin{equation}
	\begin{split}
	    \frac{d\tilde{N}_i}{d\tilde{t}} &=\tilde{\lambda} + \frac{\tilde{r}_i}{\tilde{K}_i}\tilde{N}_i\left(\tilde{K}_i - \tilde{N}_i - \sum_{j\neq i} \tilde{\alpha}_{ij} \tilde{N}_j\right)\\
	    &\phantom{=} + \sqrt{2\tilde{D}\tilde{N}_i}\eta_i(\tilde{t}) 
	\end{split}\label{eq:LV_equation}
	\end{equation}
    In this community, a species can increase in abundance by immigrating from an external species pool at a rate $\tilde{\lambda}$, or by reproducing at a rate $\tilde{r}_i$.  However, their population sizes cannot grow indefinitely due to intraspecies and interpecies competition for limited resources. In the absence of other species, the ecosystem can only sustain a finite number of individuals and this abundance ceiling is enforced by the carrying capacity $\tilde{K}_i$ and is manifested in a quadratic density-dependent death term. As for competition between different species, we make a simplifying assumption that resource dynamics occur much faster than population dynamics. As a result, interactions between species $i$ and $j$ are effectively modeled by a bilinear interaction term with a symmetric interaction matrix $\{\tilde{\alpha}_{ij}\}$. The final term in \eqref{eq:LV_equation} represents the demographic fluctuations. The noise $\eta_i(t)$ is modeled by a Wiener process with the $\langle \eta_i(t)\rangle = 0$ and correlation $\langle \eta_i(t) \eta_j(t')\rangle = \delta_{ij}\delta(t-t')$, where $\langle \ldots \rangle$ represents the long time average of a quantity. This multiplicative demographic noise is interpreted according to the Ito convention. All parameters and variables in \eqref{eq:LV_equation}, along with their definitions, are summarized in Table \ref{tab1:param_defs}

        \begin{table*}[t!] 
    \caption{\label{tab1:param_defs} Definition of parameters and variables in the stochastic Lotka-Volterra model and their corresponding dimensionless versions}
    \begin{ruledtabular}
    \begin{tabular}{clp{8cm}}
    \textrm{Parameter/Variable}& 
    \textrm{Dimensionless parameter/variable} &
    \textrm{Definition [with units]}\\
    \colrule
    $\tilde{N}_i$ & $N_i = \tilde{N}_i/\tilde{K}$ & Population size of species $i$ [abundance]\\
    $\tilde{t}$ & $t = \tilde{t}\tilde{r}_i \tilde{K}/\tilde{K}_i$ & Time in generations [time]\\
    $\tilde{r}_i$ & $r_i = \tilde{r}_i \tilde{K}/\tilde{K_i} = \text{const.}$ & Intrinsic growth rate of species $i$ [1/time]\\
    $\tilde{K}_i$ & $K_i = \tilde{K_i}/\tilde{K}$ & Intrinsic carrying capacity of species $i$ [abundance]\\
    $\tilde{K}$ & $K = 1$ & Average carrying capacity of each species [abundance]\\
    $\tilde{\sigma}_K$ & $\sigma_K = \tilde{\sigma}_K/\tilde{K}$& Standard deviation of carrying capacity [abundance]\\
    $\tilde{\alpha}_{ij}$ & $\alpha_{ij} = \tilde{\alpha}_{ij}$& Interaction strength between species $i$ and $j$ \\
    $\tilde{\mu}$ & $\mu = \tilde{\mu} $ & Average interaction strength over all pairs of species\\
    $\tilde{\sigma}$& $\sigma = \tilde{\sigma}$ & Standard deviation in interaction strengths\\
    $\tilde{D}$& $D = \tilde{D}/\tilde{K}$ & Noise strength [abundance]\\
    $\tilde{\lambda}$ & $\lambda = (\tilde{\lambda}/\tilde{K})(\tilde{K}_i/(\tilde{r}_i\tilde{K})$ & Immigration rate [abundance/time]\\
    $\eta_i$ & & Gaussian white noise with mean zero and variance one [1/time]\\
    \end{tabular}
    \end{ruledtabular}
    \end{table*}

	To incorporate species diversity and niche differences into the model, we construct a heterogeneous random community. The carrying capacity $\tilde{K}_i$ is drawn randomly from the Gaussian distribution, $P(\tilde{K}_i) \sim \mathcal{N}(\tilde{K},\tilde{\sigma}_K)$, with $\tilde{K}$ and $\tilde{\sigma}^2_K$ being the mean and variance in carrying capacity, respectively. Similarly, we pick the interaction strengths $\tilde{\alpha}_{ij}$ from another normal distribution $P(\tilde{\alpha}_{ij}) \sim \mathcal{N}(\tilde{\mu}/S, \tilde{ \sigma}/\sqrt{S})$, where $\tilde{\mu}/S$ and $\tilde{\sigma}^2/S$ represent the mean and variance in interaction strength. These random species parameters are fixed for a given community, and are hence quenched random variables. The $1/S$ scaling in the statistics of the interaction strengths $\tilde{\alpha}_{ij}$ are important because it ensures that the population neither grows without bound nor collapses too quickly. Hence, the ecosystem has an infinite species limit that would enable a statistical physics analysis.
	
	Including this population noise term is vital to accurately model ecosystem dynamics as growth, death, and interactions with biotic and abiotic environments are inherently stochastic processes. As such, the dynamical equations must reflect this randomness. Independent populations controlled by a constant average birth and death rate features fluctuations that follow Poisson statistics, scaling as $\sqrt{\tilde{N}_i}$. In reality, the presence of density dependent death, interspecific interactions, and a fluctuating environment affect the population noise as well. A quantitative prediction of ecological dynamics should include these corrections to the multiplicative noise, however, numerical studies have suggested that the qualitative phase diagram is insensitive to the precise details of the noise term \cite{fisher_transition_2014}. Thus, it suffices to model the population fluctuations of each species by an independent Wiener process with a standard deviation, $\sqrt{N_i}$, modified by an effective noise strength parameter, $D$. In this model of noise, higher levels of environmental stochasticity would translate to larger values of $D$. In addition to being a reasonable noise model of complex ecosystems with both neutral and non-neutral processes, it is one of the few analytically tractable models to solve.

    \section{Results}

    Before solving the model for the macro-ecological properties, we can reparameterize the Lotka-Volterrra equation in $\eqref{eq:LV_equation}$. Not only does this simplify the equation into a more general form, but it also reduces the space of parameters we need to explore and allows us to identify the effective parameters that govern the overall dynamics. 
    
    For example, instead of following the temporal dynamics of the absolute population size, we only need to measure the population of each species with respect to the average carrying capacity, i.e. $N_i = \tilde{N}_i/\tilde{K}$. In addition to normalizing the abundances, we make a simplifying assumption that $r_i = \tilde{r_i}\tilde{K}/\tilde{K}_i$ is a constant for all species. This allows us to rescale time for all species by $t = \tilde{t} \tilde{r}_i\tilde{K}/\tilde{K}_i$ and . As a consequence of reparameterizing $\tilde{N}_i$ and $\tilde{t}$, we rescale the noise term and define a new dimensionless parameter $D = (\tilde{D}/\tilde{K})(\tilde{K}_i/(\tilde{r}_i\tilde{K}))^2$. Since the second term in the parentheses is a constant, then the relevant scaling parameter for the noise is $\tilde{D}/\tilde{K}$. Hence, the new dimensionless form of the stochastic LV equation is 
\begin{align}
    \frac{dN_i}{dt} = \lambda + N_i \left( K_i - N_i - \sum_{j\neq{i}}\alpha_{ij}N_j\right) + \sqrt{2DN_i}\eta_i(t) \label{eq:reduced_LV}
\end{align}
All conversions between eq. \eqref{eq:LV_equation} and the dimensionless eq. \eqref{eq:reduced_LV} are summarized in the second column of Table \ref{tab1:param_defs}.

For a given set of macro-ecological parameters, we can numerically solve \eqref{eq:reduced_LV} by first drawing the interactions and carrying capacities from their corresponding distributions, and then integrating the system of stochastic differential equations using the Milstein method \cite{fisher_transition_2014}. This yields the population trajectory of each species over time, from which we can compute the equilibrium population statistics. Under certain parameter regimes, the simulation might not produce an SAD that is representative of the ecosystem. There can be large variations between different simulations of the same ecosystem because the population dynamics is noisy and the community can have multiple equilibrium solutions. Hence, averaging over multiple replicates of the ecosystem is required to obtain a more accurate species abundance distribution. The computational cost of running these replicated simulations makes it challenging to explore the vast parameter space and determine the boundaries that separate the ecosystem with different SAD behaviors. As such, we seek an analytical solution to eq. \eqref{eq:reduced_LV}.

\floatsetup[figure]{style=plain,subcapbesideposition=top}
\begin{figure*}[t!]
    \centering
    \sidesubfloat[]{
        \includegraphics[width=0.2\textwidth]{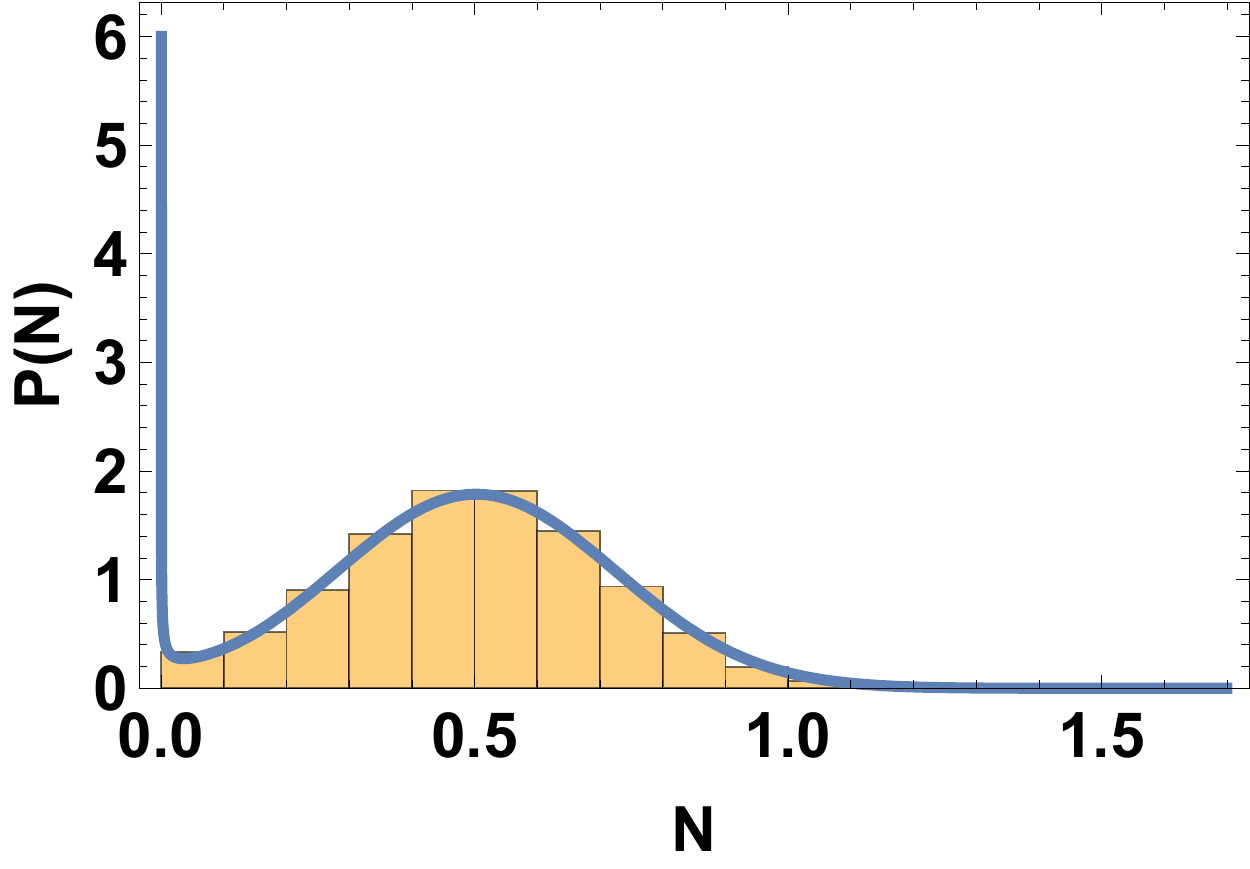}
        \includegraphics[width=0.2\textwidth]{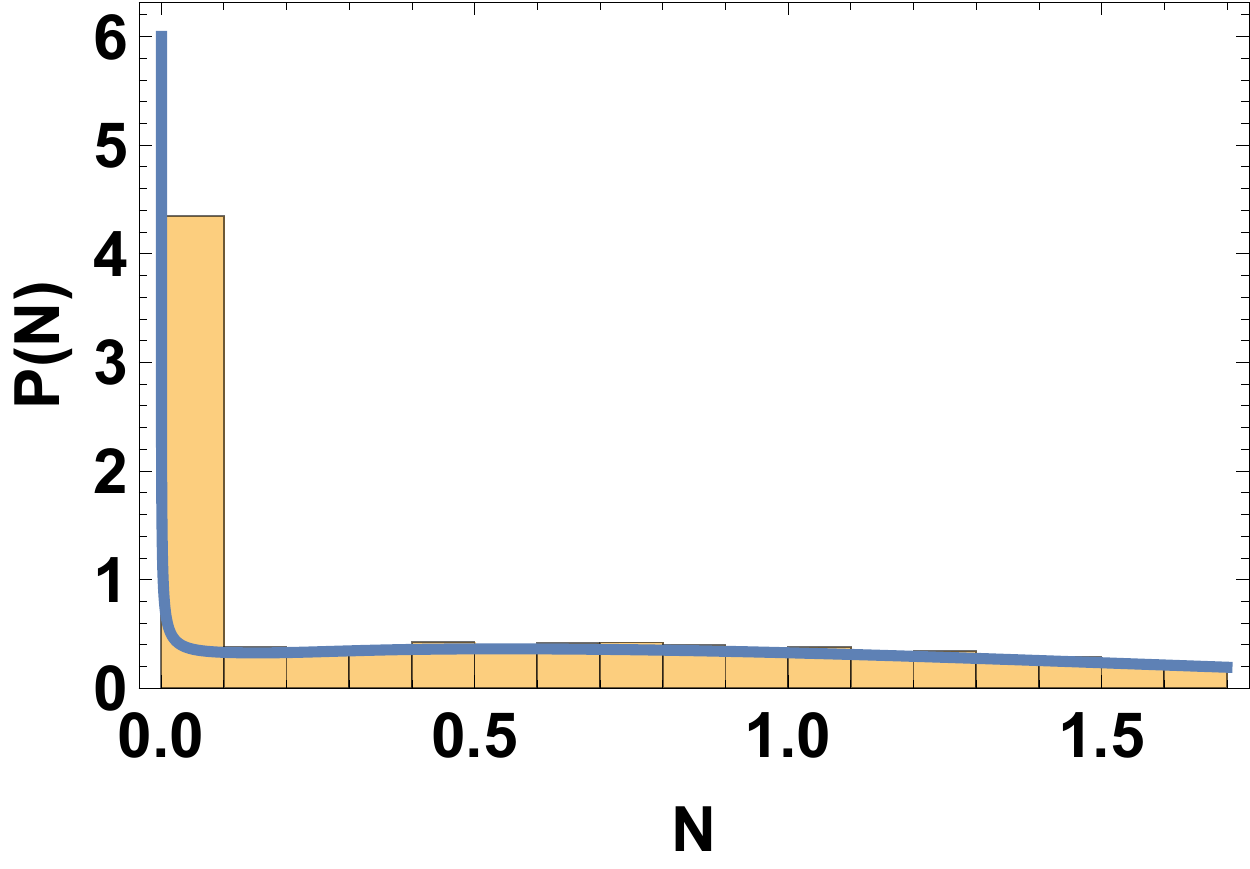}
        \includegraphics[width=0.2\textwidth]{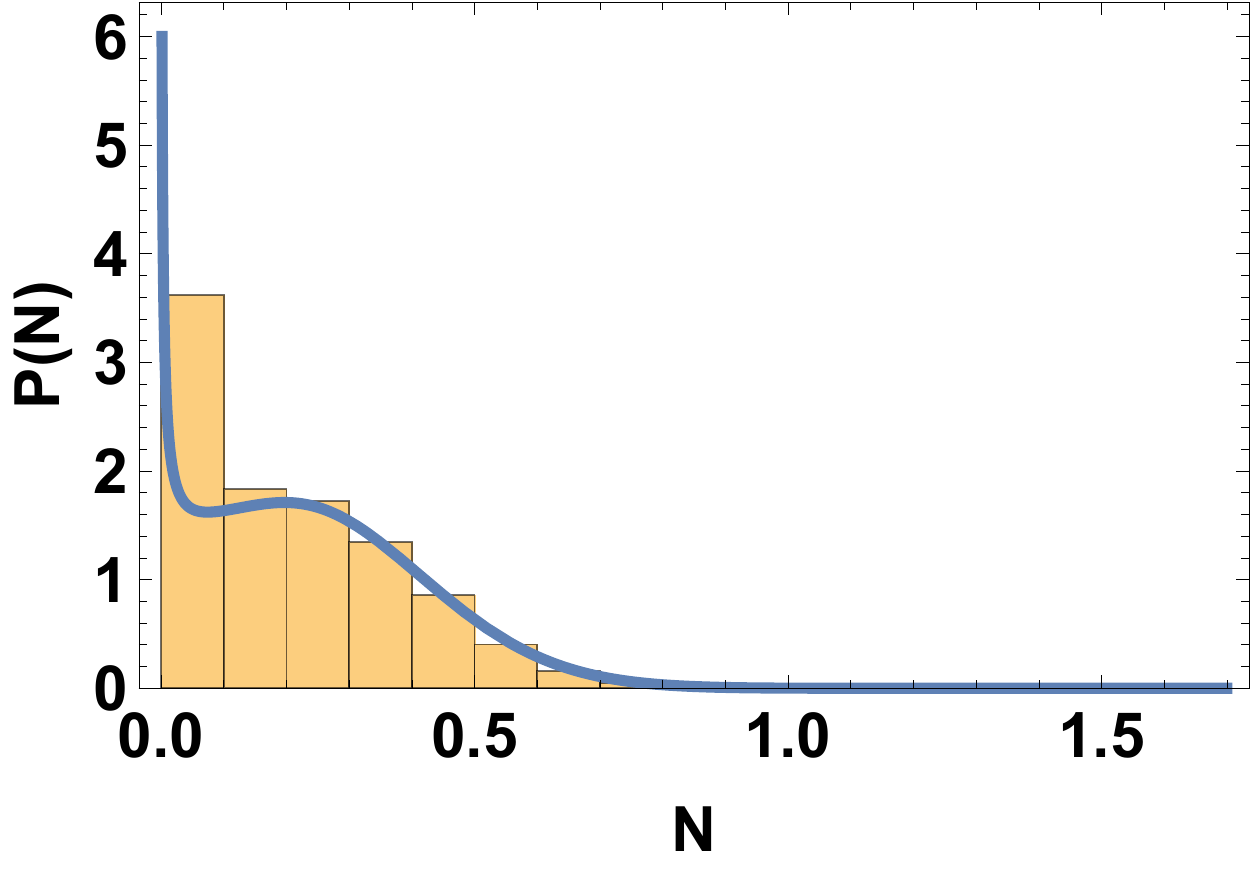}
        \includegraphics[width=0.2\textwidth]{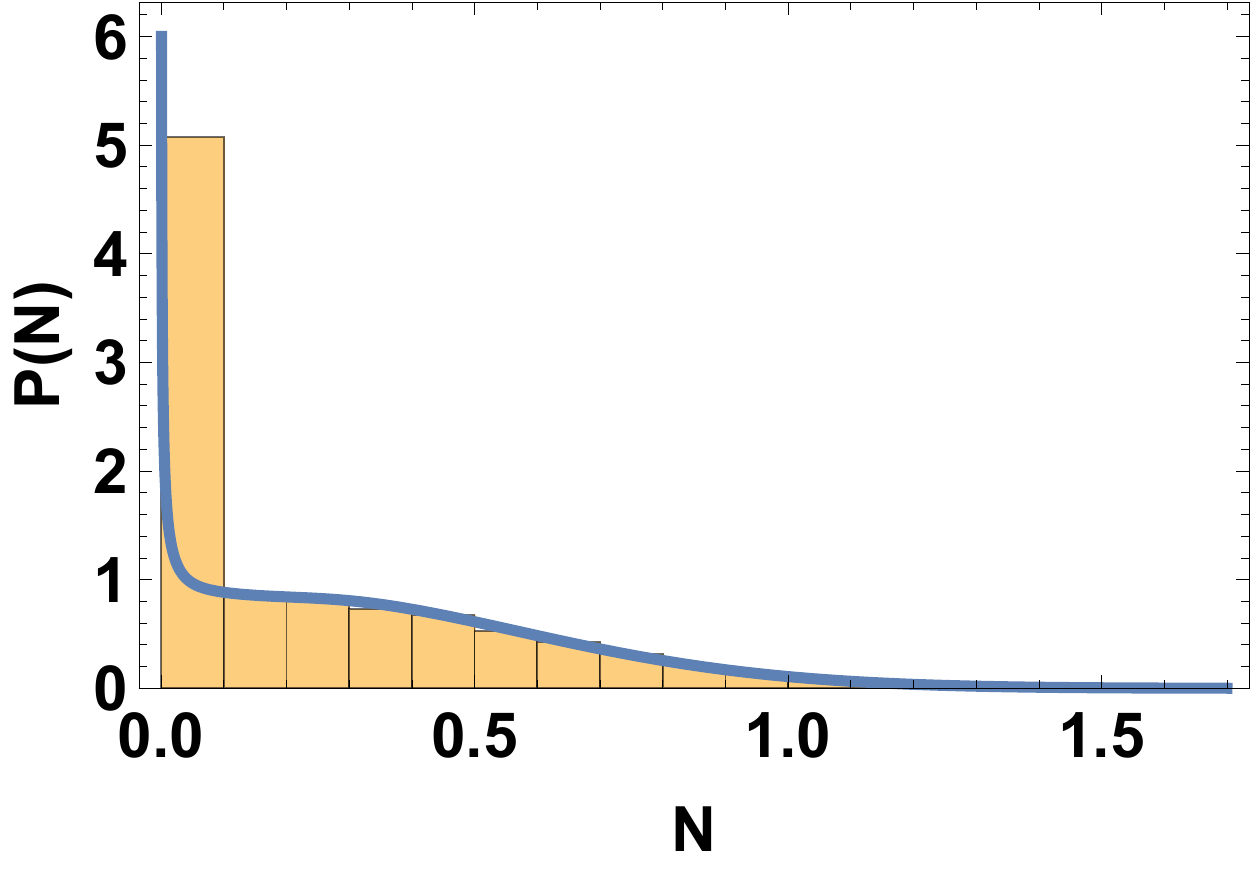}
    }\\
    \sidesubfloat[]{
        \includegraphics[width=0.2\textwidth]{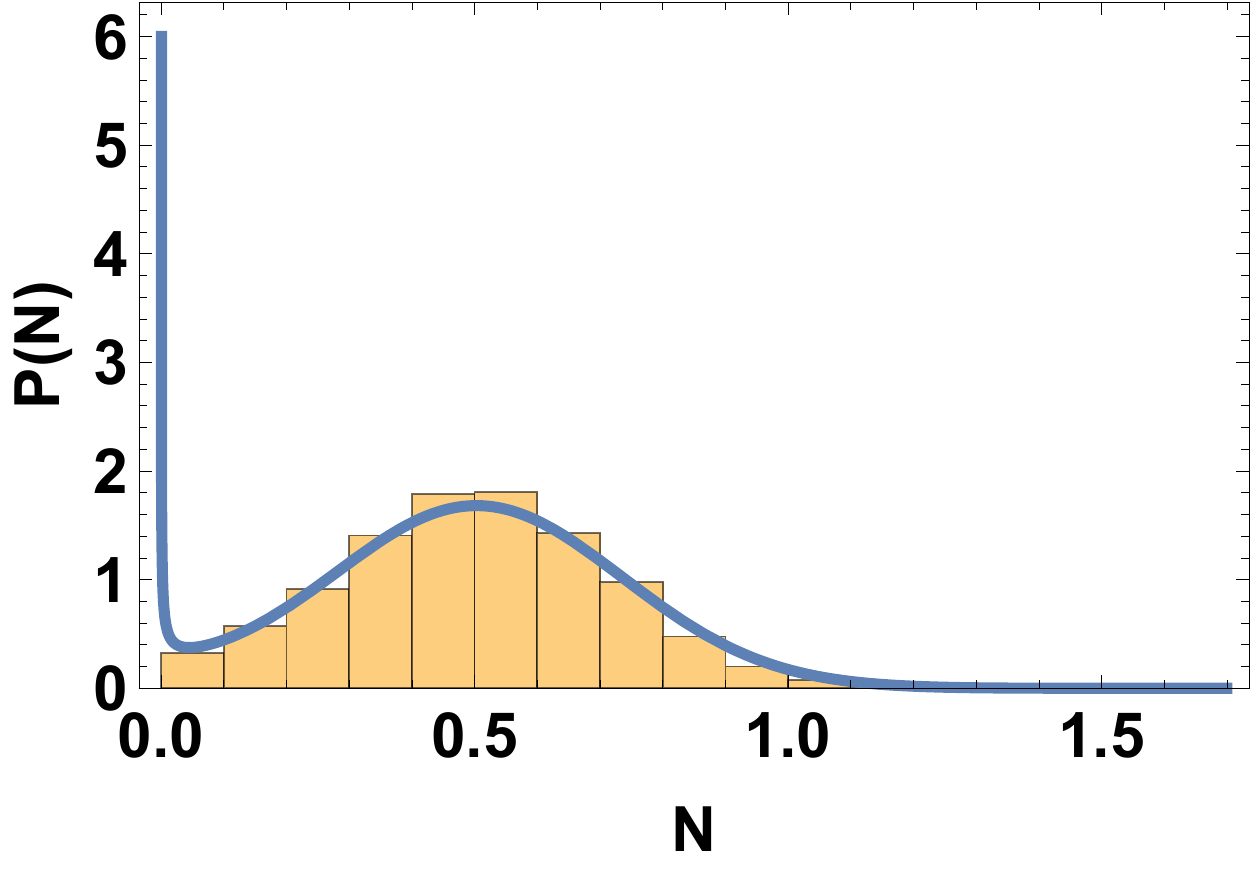}
        \includegraphics[width=0.2\textwidth]{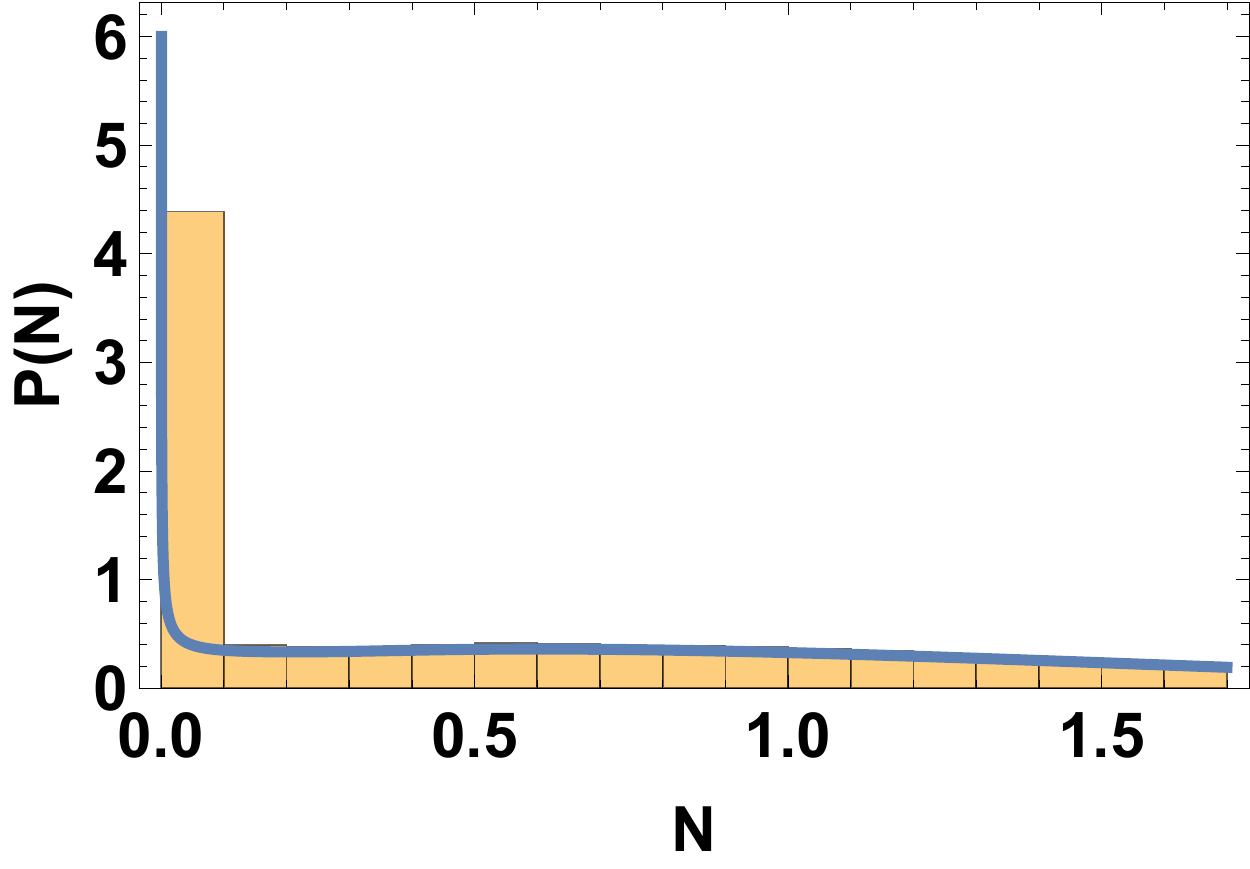}
        \includegraphics[width=0.2\textwidth]{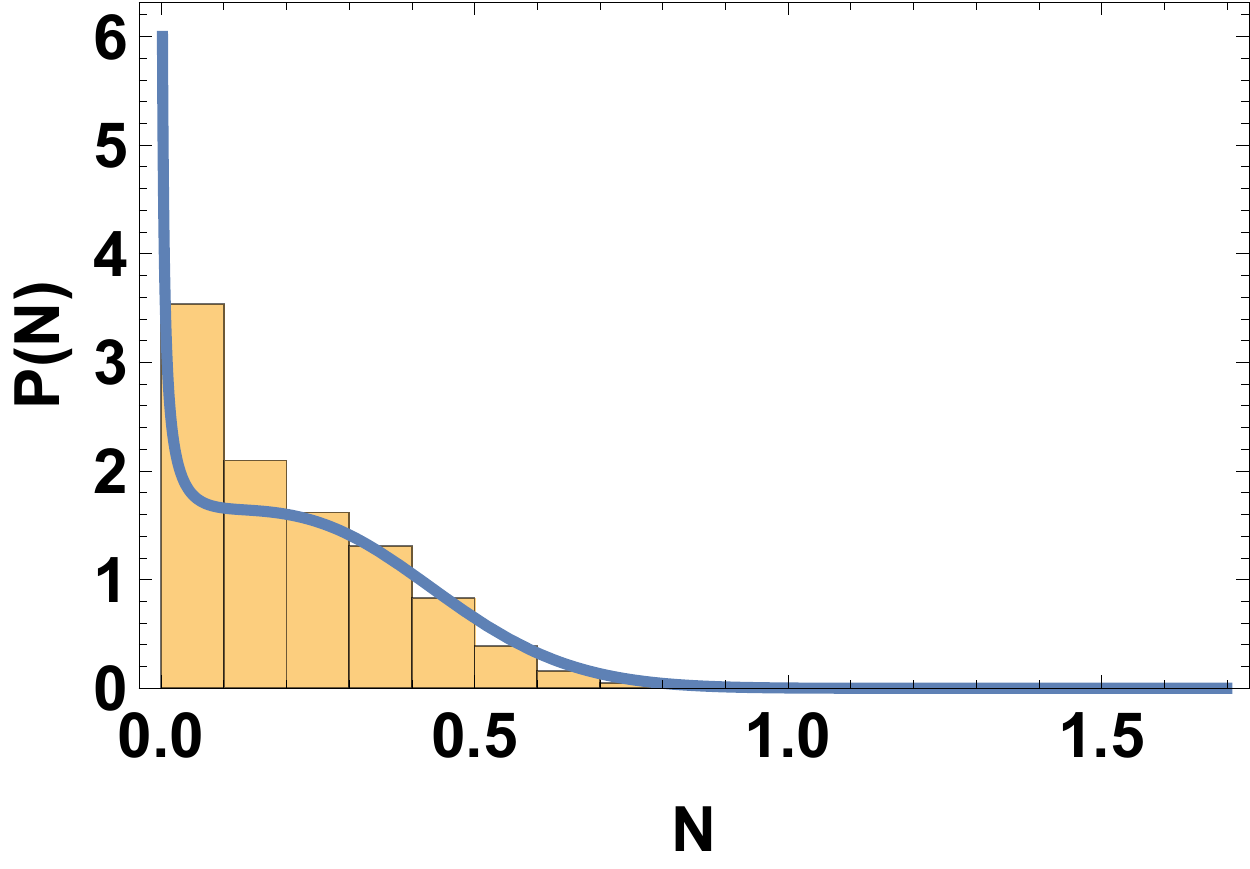}
        \includegraphics[width=0.2\textwidth]{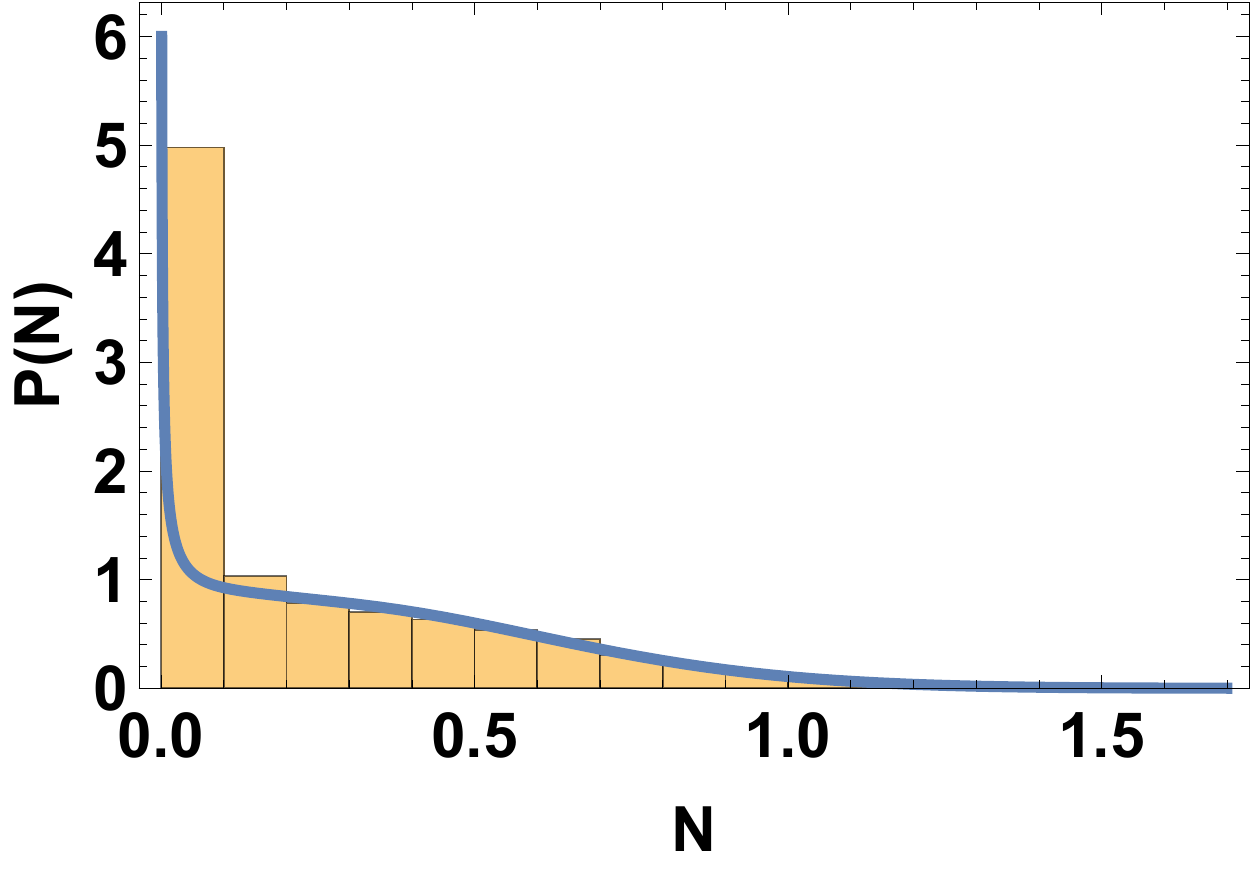}
    }\\
    \sidesubfloat[]{
        \includegraphics[width=0.2\textwidth]{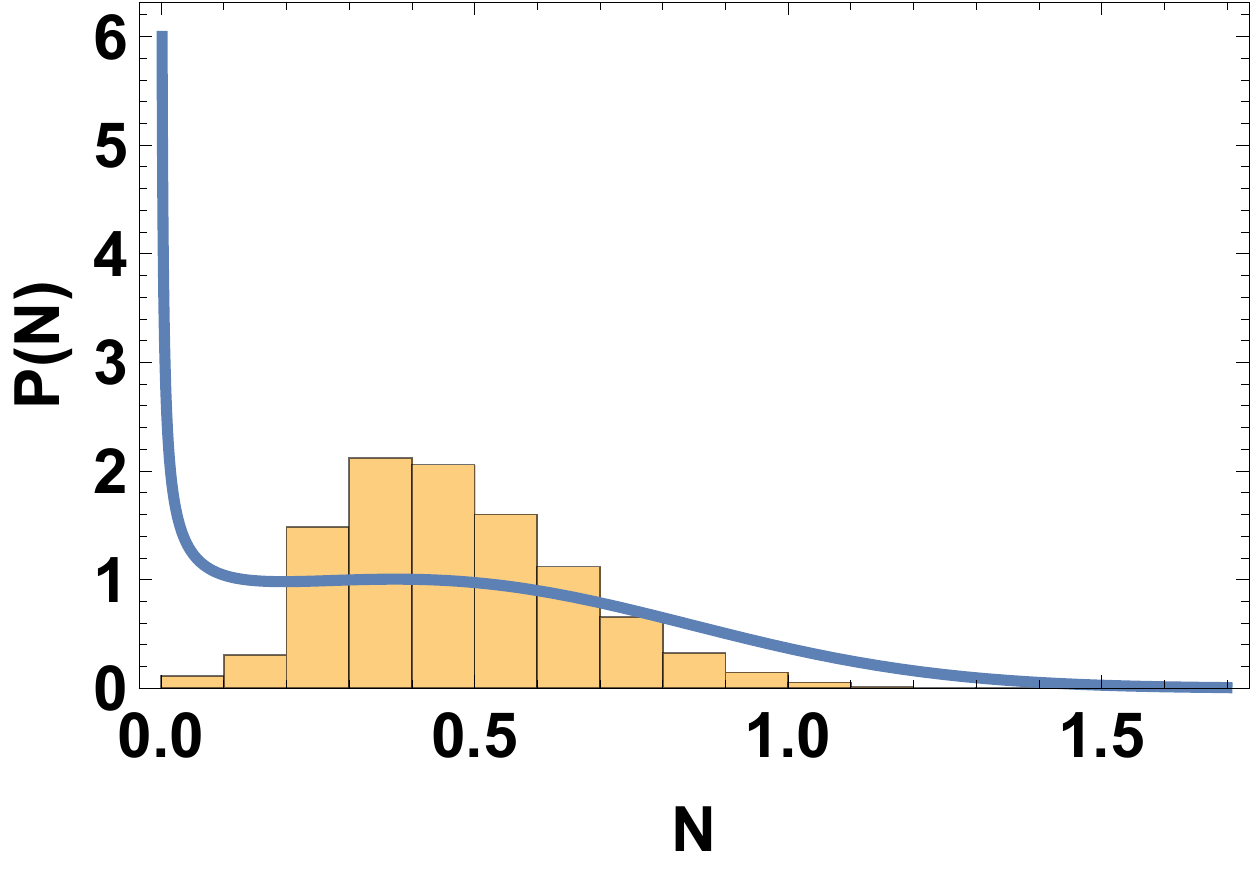}
        \includegraphics[width=0.2\textwidth]{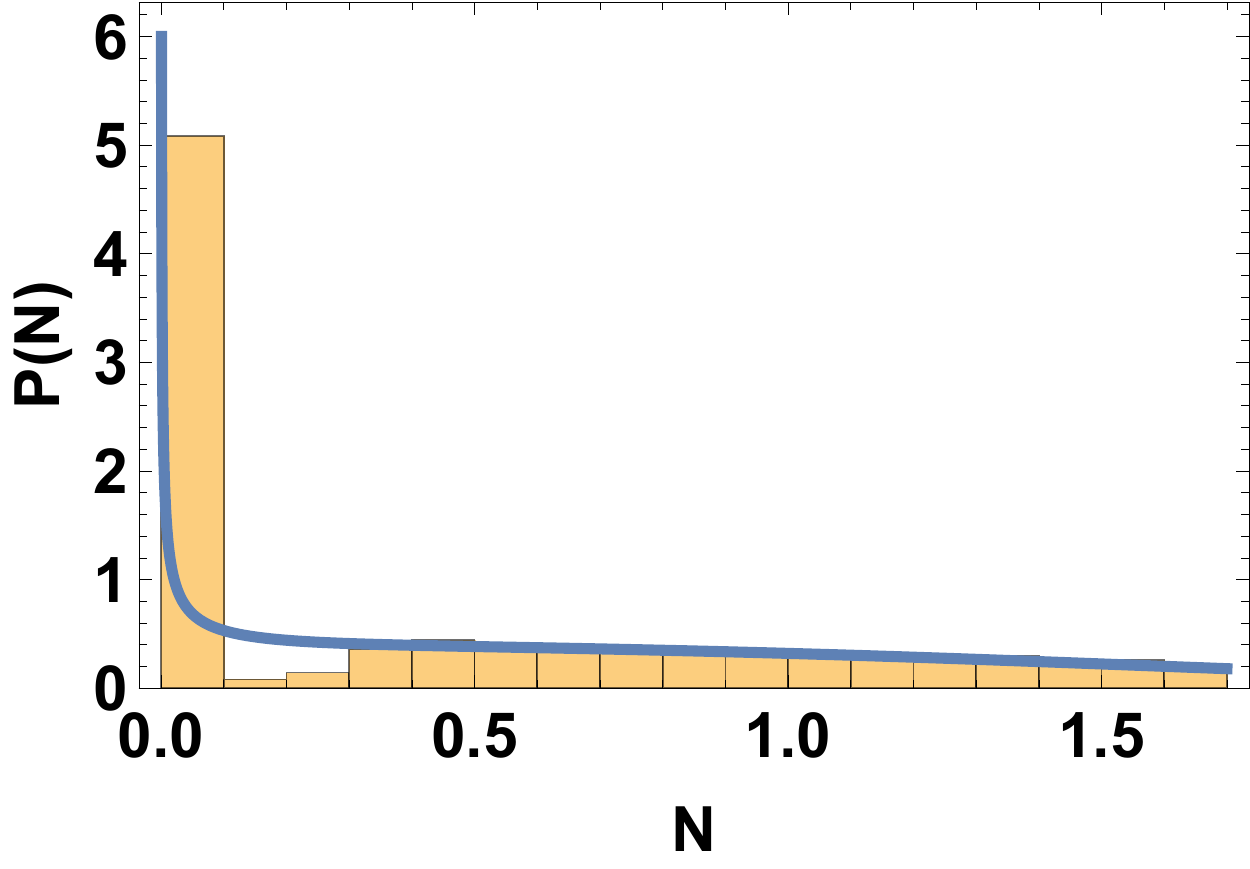}
        \includegraphics[width=0.2\textwidth]{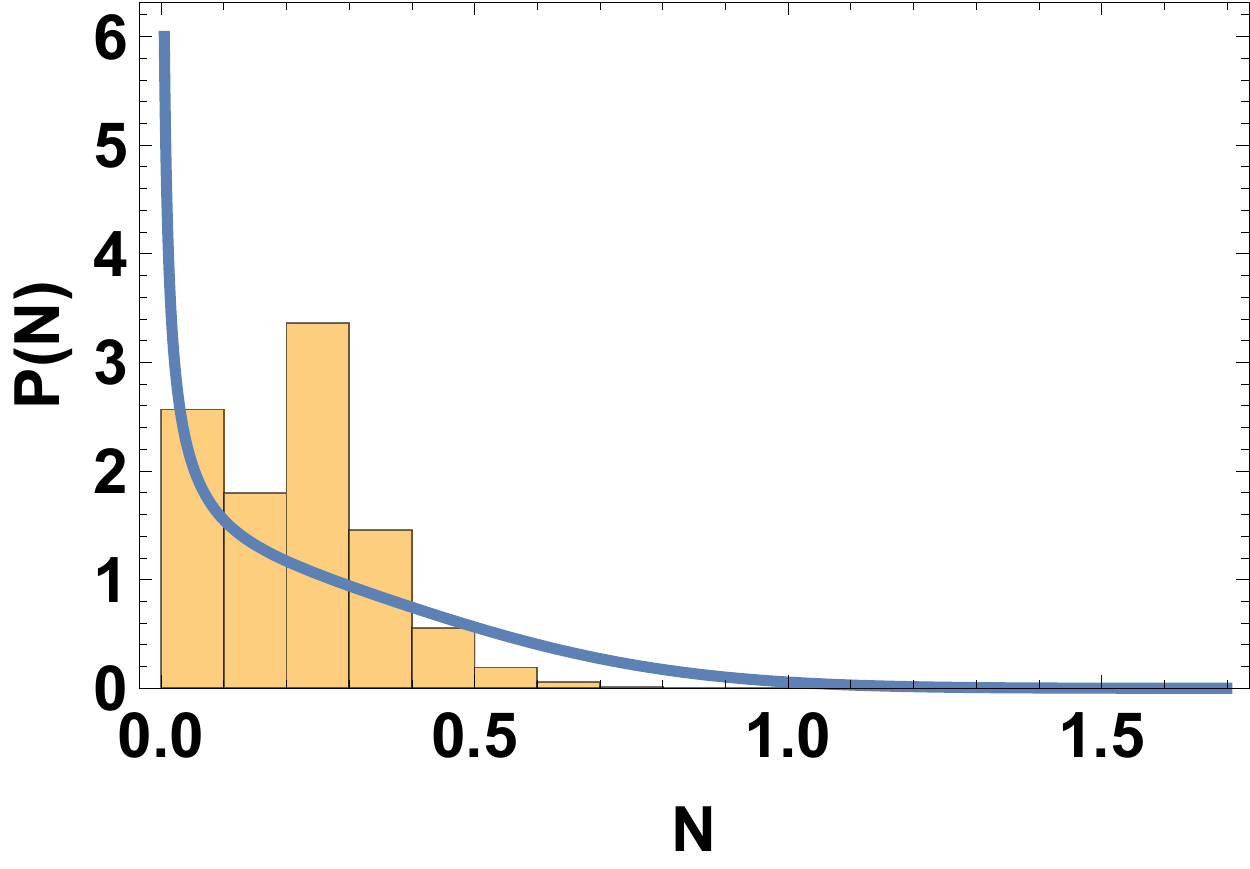}
        \includegraphics[width=0.2\textwidth]{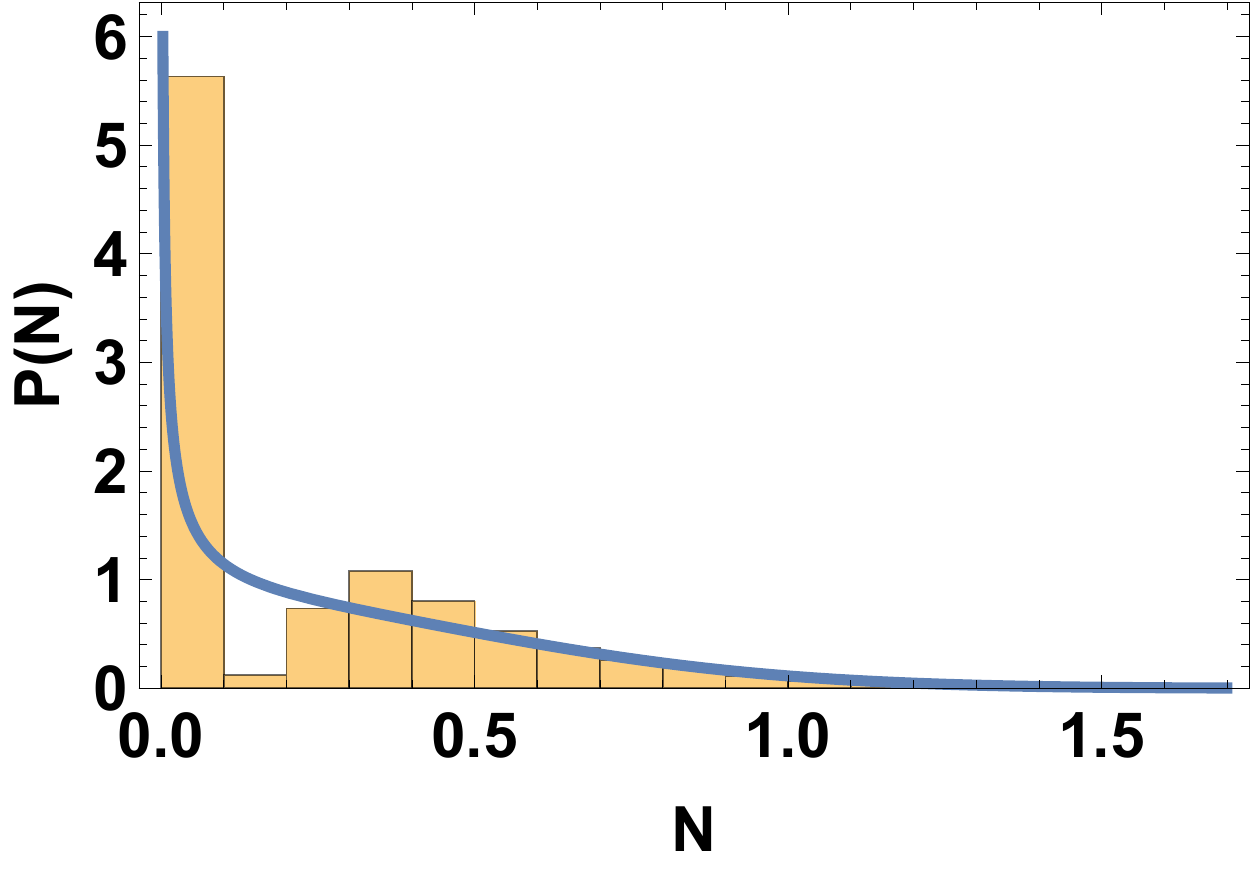}
    }\\
    \sidesubfloat[]{
        \includegraphics[width=0.2\textwidth]{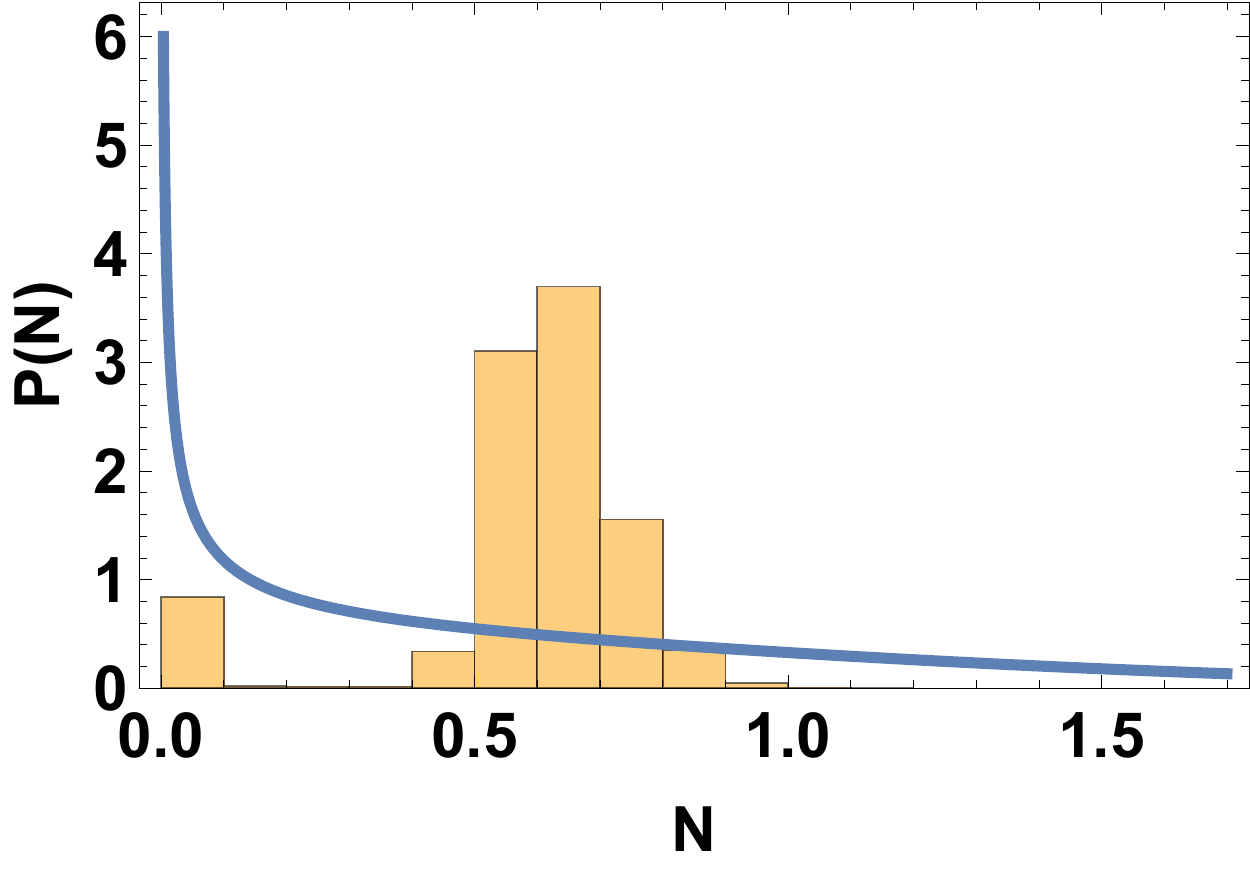}
        \includegraphics[width=0.2\textwidth]{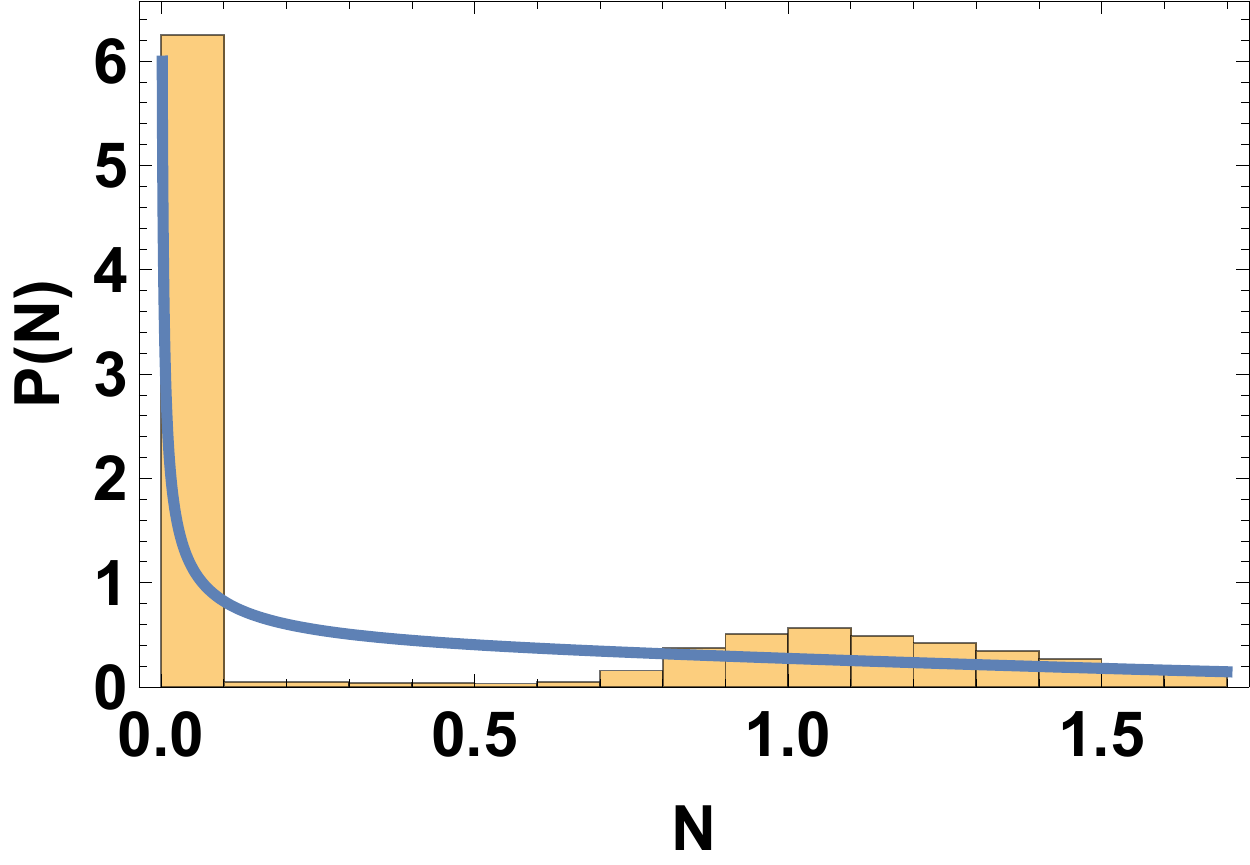}
        \includegraphics[width=0.2\textwidth]{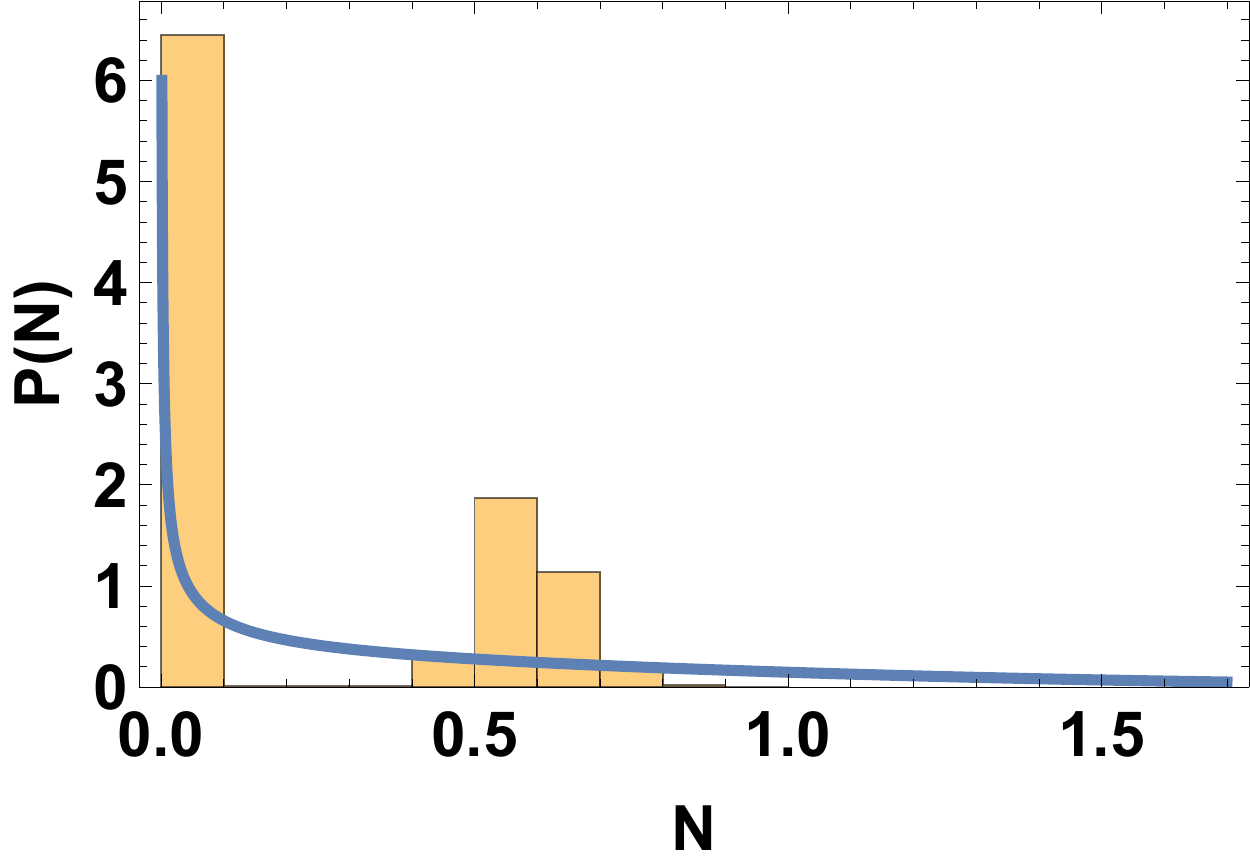}
        \includegraphics[width=0.2\textwidth]{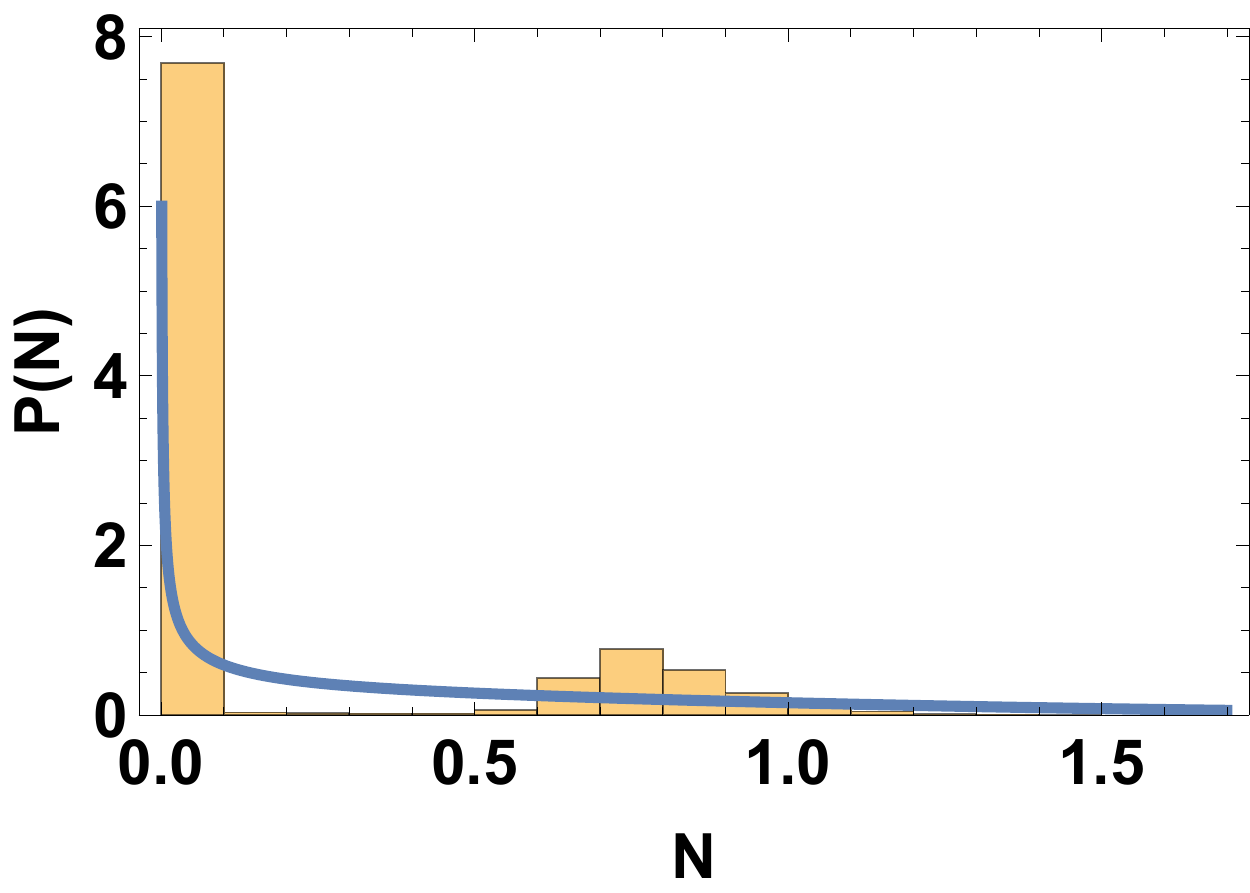}
    }
    \caption{ Marginal distributions of species abundance. Using population trajectories from integrating the stochastic Lotka-Volterra equations, we extracted the population mean of each species to construct a probability density histogram of the species abundances (orange bars). Each histogram consists of 30 ecosystem replicates each with 1000 species. Additionally, we solved for the theoretical population statistics using the iterative scheme and used these quantities to calculate the marginal distribution $P(N)$ (blue line). For our numerical computations, we set $(K,\sigma_K) = (1,0.2)$. Each row of the figure corresponds to a different level of noise: (a) $D = 0.005$, (b) $D = 0.01$, (c) $D = 0.1$ , and (d) $D = 1.0$. Each column represents the lower left (first column), upper left (second column), lower right (third column), and upper right (fourth column) corners of the $\mu$-$\sigma$ phase diagram, respectively. First column: $(\mu,\sigma) = (1.0,0.1)$. Second column: $(\mu,\sigma)  = (1.0,0.6)$. Third column: $(\mu,\sigma) = (4.0,0.1)$. Fourth column: $(\mu,\sigma) = (4.0,0.6)$. The distributions from the Milstein method and iterative scheme tend to agree for small values of $D$, but tend to be different as we increase the noise level.Furthermore, as we move toward the lower right diagrams, noise increases and the variability in interactions decreases, corresponding to the neutral-like regime. The theoretical probability distribution of species abundances (blue) start to look more generic with a rapid decay towards zero as $N$ increases and a divergence at $N = 0$. This is quite reminiscent of the Fisher log-series distribution commonly found in neutral ecosystems except with a continuous $N$ rather than a discrete one.} 
    \label{fig:marginal_dist_comp}
\end{figure*}

\subsection{Analytical Solution}

To solve eq.\eqref{eq:reduced_LV} analytically, we transform it into the corresponding Fokker-Planck equation (FPE).
In exchange for the exact population trajectory of each species, we gain access to the distribution of population sizes over time. As we are only concerned with the ecosystem at long times, we only need to solve for the stationary joint distribution of abundances of all species, $P_S(\{N_i\})$. Since the interaction matrix $\alpha_{ij}$ is symmetric, we can find an exact solution that resembles the Boltzmann distribution from statistical physics.

Even with the stationary joint distribution, extracting population statistics for a representative species from $P_S(\{N_i\})$ remains challenging due to the quenched random variables $K_i$ and $\alpha_{ij}$. The joint distribution for all species is also rather uninformative due to sheer number of species represented in the distribution. Instead, distributions and statistics of a representative species of the community are preferred. Long-time average population statistics such as the mean $\langle N_i\rangle$ and second moment $\langle N_i^2\rangle$ in abundance generally depend on the exact values of carrying capacities $K_i$, interactions $\alpha_{ij}$, and the abundance of all other species. However, in the limit $S\rightarrow \infty$ where the community is highly diverse, macro-ecological statistics simplify and only depend on the bulk parameters $D,K,\sigma_K,\mu,$ and $\sigma$. To handle the quenched random variables and obtain the statistics of a representative species, we draw on techniques from the physics of disordered systems.

Naively, one could treat the interaction field term $h_i = \sum_{j=1}\alpha_{ij}N_j$ as a sum of many independent random variables, and therefore be approximated by its mean via the law of large numbers. This is the naive mean field approach and it yields the incorrect population statistics. Let us consider the population dynamics of species $i$. Changes in the abundance $N_i$ ripples throughout the community, generating fluctuations in the population sizes $N_j$ of other species and hence $h_i$. In turn, this change in abundances of other species in the community affects the population of species $i$. It turns out that the fluctuations in $h_i$ are of the same order as its mean, and therefore cannot be neglected \cite{advani_statistical_2018,nishimori_statistical_2001,mezard_spin_1987,del_ferraro_cavity_2014}. To truly capture these reactions and correlations in the community, we rely on the replica symmetric cavity method from the physics of disordered systems. . 

The cavity method starts by introducing a new species $0$ into the community. In the thermodynamic limit $S\rightarrow \infty$, the community with $S+1$ species exhibits exhibits the same statistical properties as the original one with $S$ species.
Now, let us define a new cavity field $h_c =  \sum_{j=1}^S \alpha_{0j}N_j$, the local field felt by new species 0 due to interactions with all of the original species. By construction, these new interaction strengths $\alpha_{0j}$ are completely uncorrelated to the abundances $N_j$. Hence, the advantage of the replica symmetric cavity method is that we can leverage the central limit theorem and treat the cavity field $h_c$ as a Gaussian random variable with distribution $P(h_c)$ instead of a quantity that depends on the original populations $\{N_j\}$. 

Now, upon replacing the interaction term with $h_c$, we can integrate over $h_c$ to obtain a marginal distribution for the new species, $P_{S+1}(N_0)$ (See Supplementary Information). From this marginal distribution, we can extract the population statistics of species 0. Given that the species pool is large, $S\rightarrow \infty$, then the statistics of species 0 are no different than any other species in the $S+1$ ecosystem. Although the invasion of the new species can cause the community to reshuffle and the equilibrium abundances of each species to change, the overarching statistics of the entire ecosystem and the species abundance distribution should remain unaltered. Namely, the marginal distribution for a representative species is 
\begin{widetext}
\begin{equation}
    P_{S+1}(N_0) = \int  \frac{Dz }{\mathcal{N}(z) N_0^{1-\beta\lambda}} \exp\left[- \frac{\beta}{2}\left( (1 - \sigma^2\beta\Delta q) N_0^2 -2(K - \mu \overline{\langle N\rangle} + \sqrt{\sigma_K^2 + q\sigma^2}z)N_0\right) \right]\label{eq:marginalDist}
\end{equation}
\end{widetext}
Here, $\overline{\langle ...\rangle}$ represents the quenched average of a population statistic, where the average is taken over long times and the ensemble of possible ecological communities, i.e. the distribution of possible $\alpha_{ij}$ and $K_i$. So, $\overline{\langle N\rangle}$, $\overline{\langle N^2\rangle}$, $q = \overline{\langle N\rangle^2}$, and $\Delta q = \overline{\langle N^2\rangle} - \overline{\langle N\rangle^2}$ represent the quenched mean population size, the quenched mean squared abundance, the quenched average of the square of the mean abundance, and the quenched variance in population size, respectively. The equations for these quantities can be derived using the cavity method and they must be computed self-consistently. Furthermore, we define $\beta = 1/D$ as the inverse ''temperature'' of the community and $\mathcal{N}(z)$ as the normalization of the distribution. Also, note that the marginal distribution in \eqref{eq:marginalDist} consists of integrating a function over a Gaussian random variable $Dz = dz~e^{-z^2/2}$, which captures the ensemble average over all possible ecosystems. The derivation of the marginal distribution is detailed in the Supplementary Information.

\subsection{Comparing Numerical to Analytical Marginal Distributions}

To verify our marginal distribution in equation \eqref{eq:marginalDist}, we compare it to the results of numerical analysis of the stochastic Lotka-Volterra equation. As shown in Fig. \ref{fig:marginal_dist_comp}, the population trajectories of 1000 species are computed in 30 replicate ecosystems (with parameters $D,K,\sigma_K,\mu,\sigma,\lambda$) and their long-time equilibrium abundances $\langle N\rangle$ are collected into a histogram and normalized to construct a probability density function. 

\begin{figure}[t!]
    \centering
    \sidesubfloat[]{\includegraphics[width=0.43\textwidth,trim=0.1cm 1cm 0.1cm 1cm, clip]{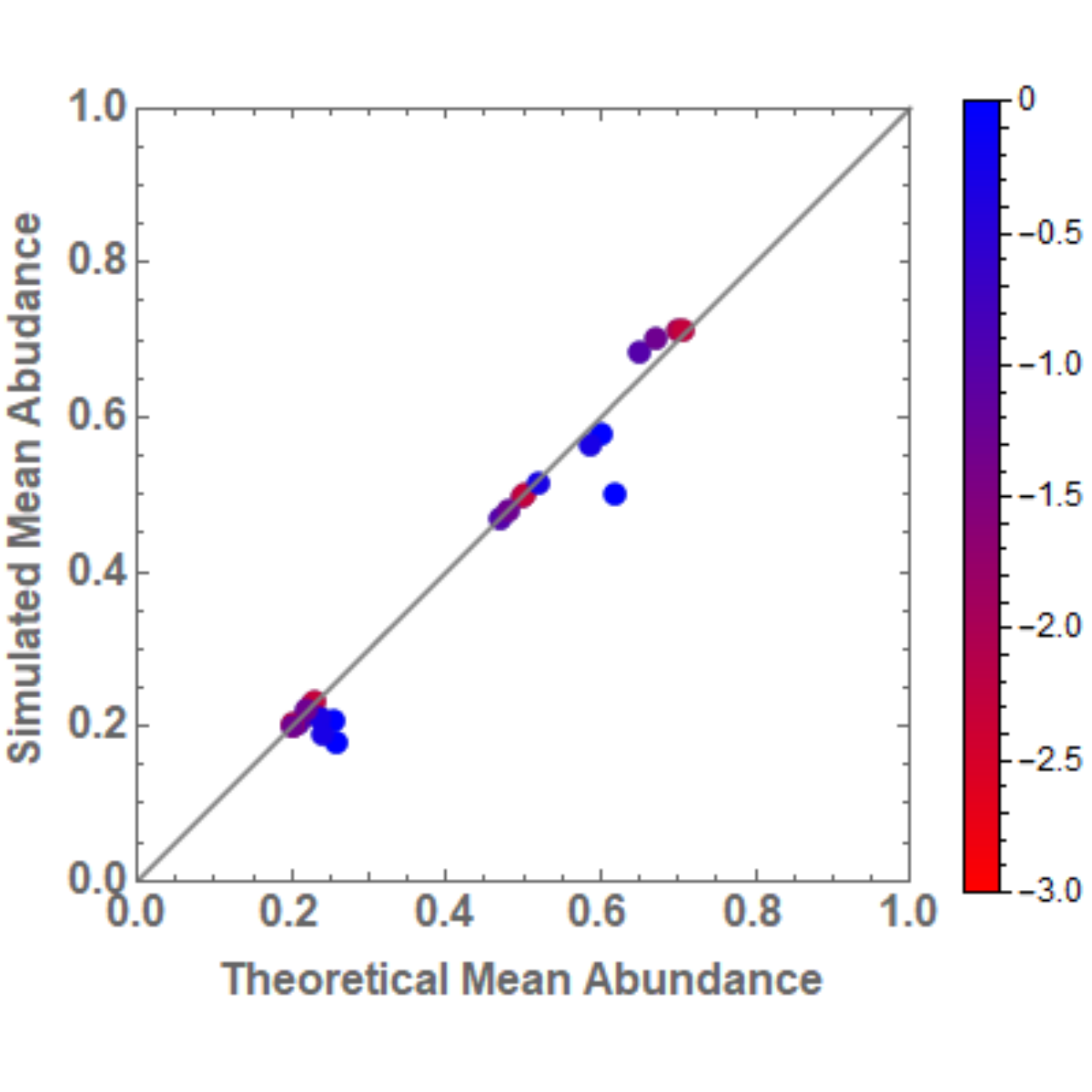}}
    \sidesubfloat[]{\includegraphics[width=0.43\textwidth,trim=0.1cm 1cm 0.1cm 1cm, clip]{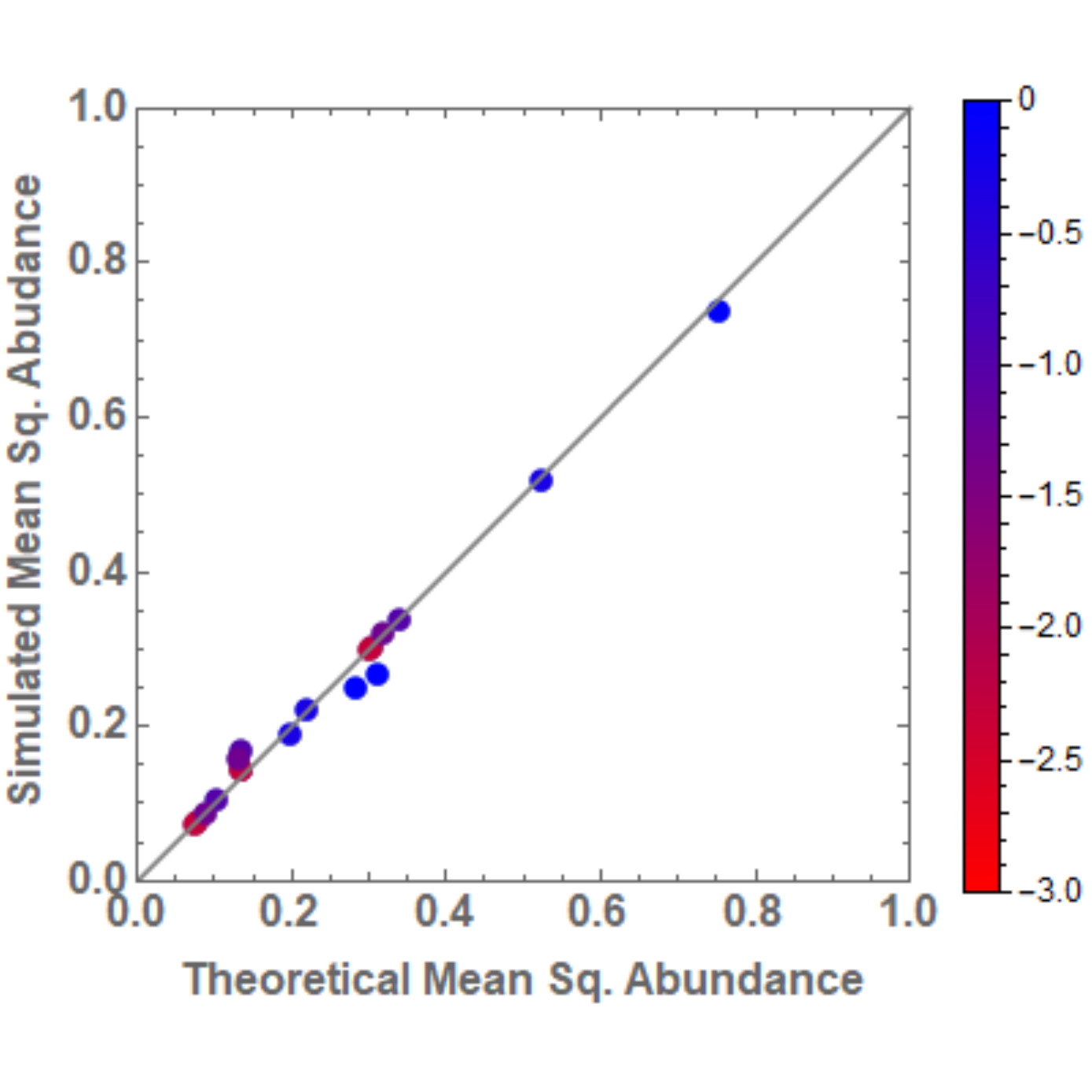}}\\
    \sidesubfloat[]{\includegraphics[width=0.43\textwidth,trim=0.1cm 1cm 0.1cm 1cm, clip]{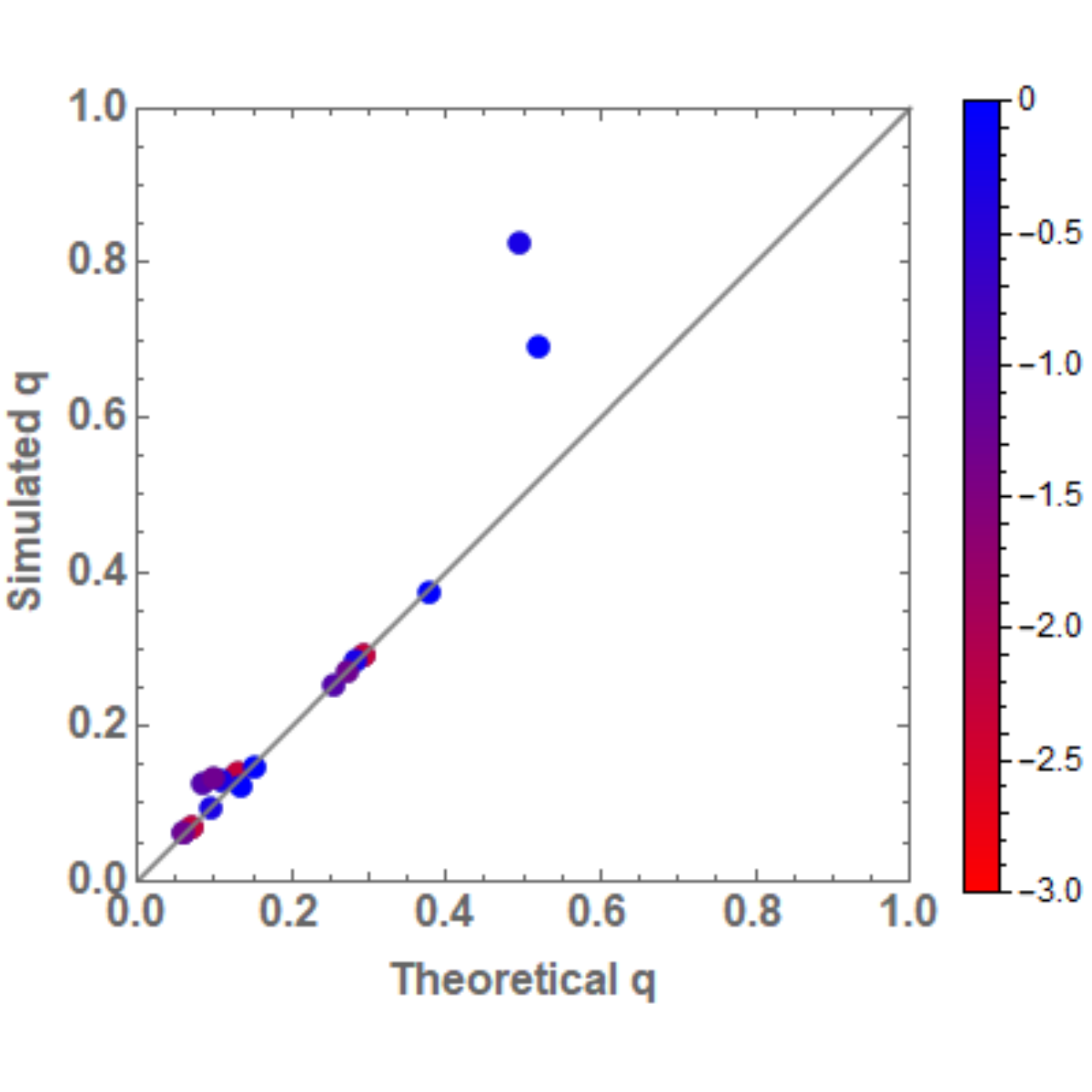}}
    \sidesubfloat[]{\includegraphics[width=0.43\textwidth,trim=0.1cm 1cm 0.1cm 1cm, clip]{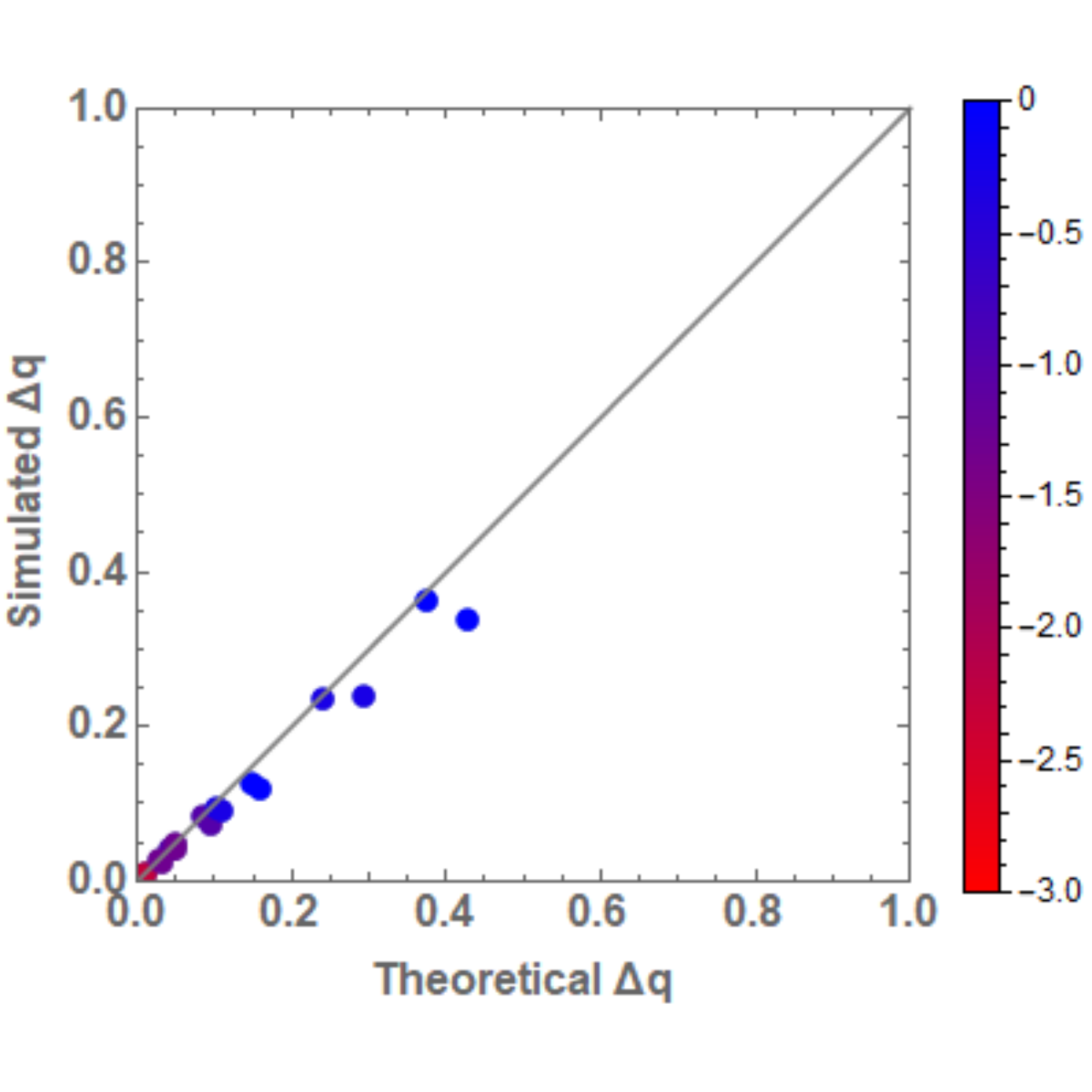}}
    \caption{A comparison between population statistics computed from solving the theoretical self-consistent equations and integrating the LV equations. Each plot corresponds to a different statistics: (a) the quenched mean abundance $\overline{\langle N\rangle}$, (b) the quenched mean squared abundance $\overline{\langle N^2\rangle}$, (c) $q = \overline{\langle N\rangle^2}$, and (d) $ \Delta q =  \overline{\langle N^2\rangle - \langle N\rangle^2}$. The quantities obtained from either method tend to agree with each other, however there are some deviations especially for cases where the noise level $D$ becomes large.}
    \label{fig:theory_sim_comp}
\end{figure}

First, for many parameter sets, the constructed histogram generally agrees with the corresponding marginal distributed computed from equation \eqref{eq:marginalDist}. When the strength of stochasticity $D$ is high, there tends to between the near-extinct species and the more abundant species. This is expected in numerical simulations as high levels of noise can quickly drive rare species to extinction. Only those with higher abundances would have a good chance of overcoming the noise and surviving. Despite some mismatch between numerical histograms and the theoretical marginal distribution, the quenched statistics $\overline{\langle N\rangle},\overline{\langle N^2\rangle},q,$ and $\Delta q$ calculated from the histogram match those computed self-consistently from the cavity method. The comparisons of the quenched statistics are shown in Fig. \ref{fig:theory_sim_comp}. 

Second, the theoretical marginal distributions tend to fall into the typical categories of species abundance distributions. As seen in Fig. \ref{fig:theory_sim_comp}, an ecosystem where noise levels $D$ are low and species are less competitive (i.e. when $\mu$ is low) tend to have SADs that have one internal mode at high populations and an integrable singularity at zero abundance. On the other hand, ecosystems with high levels of stochasticity and stronger competition exhibit SADs that resemble the Fisher log-series with a monotonic decay in probability density as abundance increases. In addition to these distributions, there are SADs with an inflection point. The parameters corresponding to these communities represent the transition point that separate the ecosystems with log-series-like SADs from those with an internal mode.

There is another transition line that is characterized by the balance of stochastic effects from demographic noise and migration. From the marginal distribution in eq.\eqref{eq:marginalDist}, when there is a net migration away from the community to other ecosystems, $\beta\lambda < 0$, then $P_{S+1}(N_0)$ contains a non-integrable singularity. Hence, the steady-state of the stochastic LV equations is a delta-function at $N_0 = 0$ with all species ultimately going extinct. If $0< \beta\lambda < 1$, then the SADs have an integrable singularity and a possible internal mode, as seen in Fig. \ref{fig:theory_sim_comp}. However, if the effects of immigration is stronger than demographic stochasticity, $\beta\lambda > 1$, then the singularity disappears. All species can be supported at least by immigration and the SAD has a peak with few species at very low and very high abundances. Although this type of SAD is not the log-normal distribution, it shares many of its general characteristics and agrees qualitatively with neutral theories of local communities.

\floatsetup[figure]{style=plain,subcapbesideposition=top}

\subsection{SAD Phase Diagram}

\begin{figure}
    \centering
    \sidesubfloat[]{\includegraphics[width=0.9\textwidth]{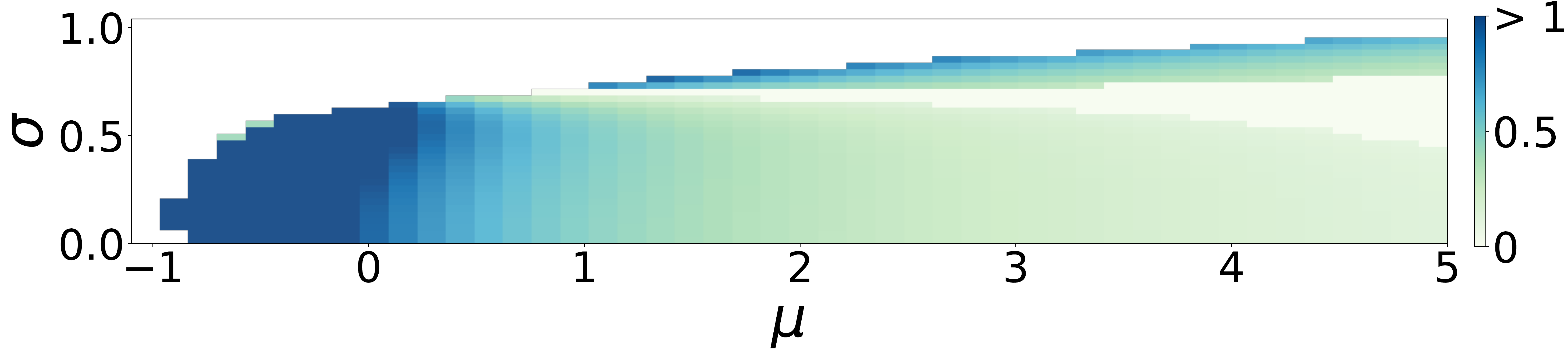}}\\
    \sidesubfloat[]{\includegraphics[width=0.9\textwidth]{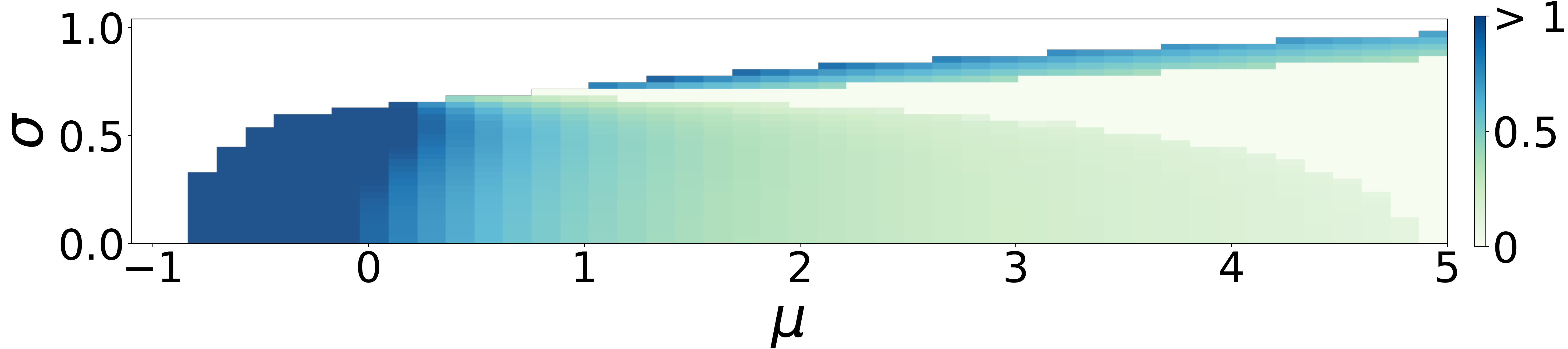}}\\
    \sidesubfloat[]{\includegraphics[width=0.9\textwidth]{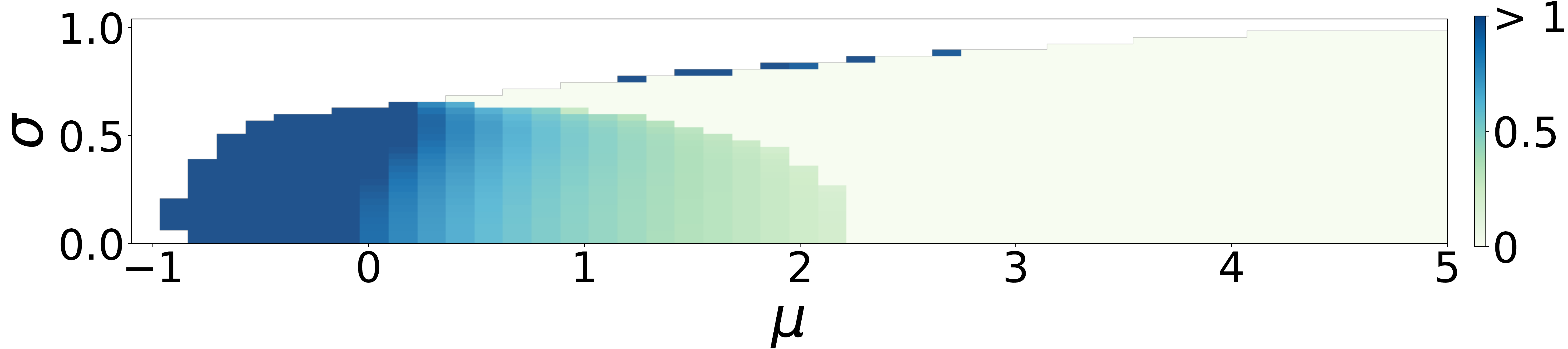}}\\
    \sidesubfloat[]{\includegraphics[width=0.9\textwidth]{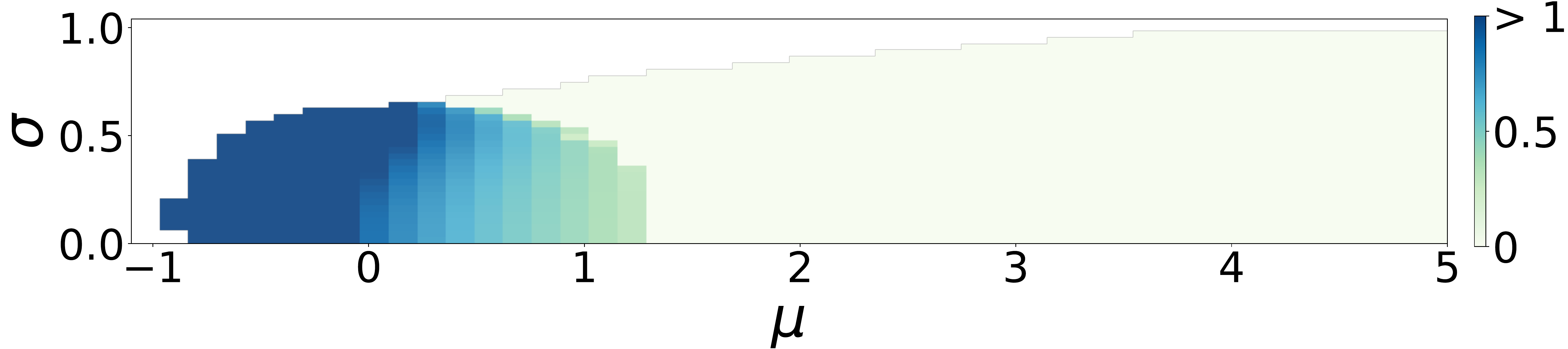}}\\
    \sidesubfloat[]{\includegraphics[width=0.9\textwidth]{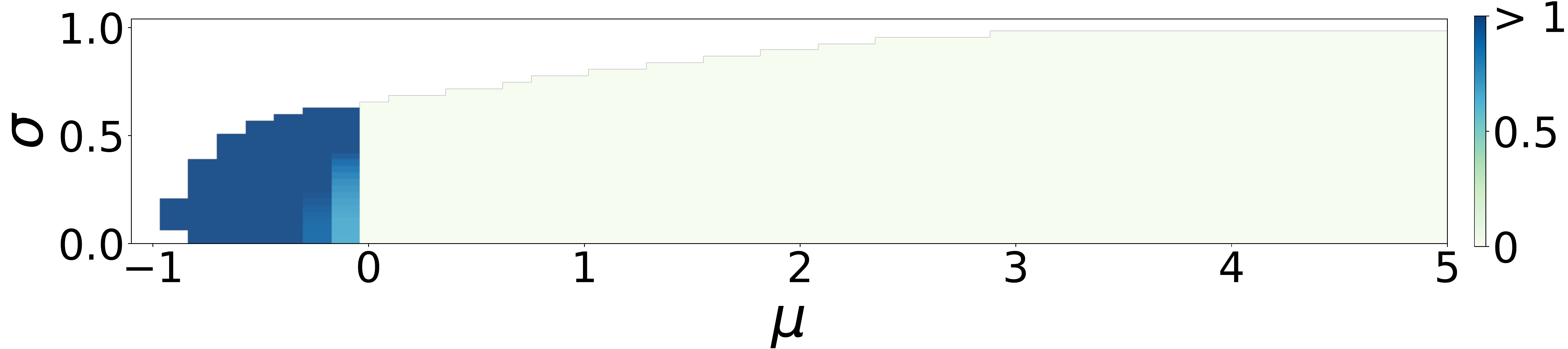}}\\
    \sidesubfloat[]{\includegraphics[width=0.9\textwidth]{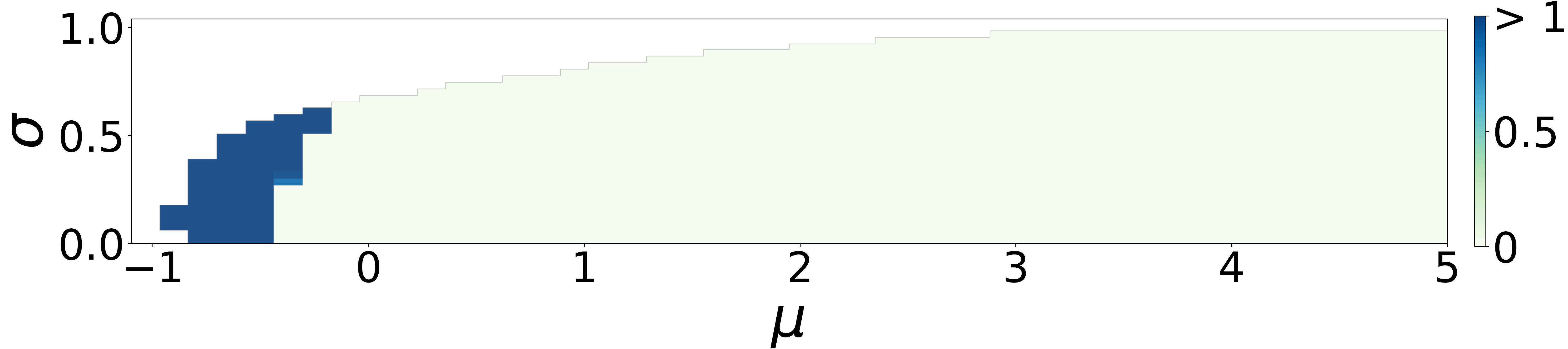}}\\
    
    \caption{Phase diagram for the location of the secondary maximum of the marginal distribution. Light green corresponds to a unimodal abundance distribution that exhibits an integrable singularity at $N=0$, which is representative of the SADs of neutral communities. Blue represents ecosystems with bimodal abundance distributions with a secondary maximum at $N_* > 0$, which is representative of SADs of communities dominated by niche differentiation. Larger values of $N_*$ is signified by a darker shade of blue. Each phase diagram corresponds to a different temperature: $D = $ (a) 0.005, (b) 0.01, (c) 0.05, (d) 0.1, (e) 0.5, (f) 1.0. At high levels of stochasticity ,the community is primarily dominated by neutral processes and the SADs exhibit unimodality. However, the community can still have a secondary local maximum if $-1< \mu < 0$, which means that cooperation at low levels can stabilize the coexistence of many species at high abundance in the face of strong stochasticity. However, at higher levels of cooperation and larger variations $\sigma$ between species, the ecosystem destabilizes and collapses with most species going extinct and few surviving. On the other hand, as the noise level $D$ decreases, it becomes increasingly difficult for cooperative communities to maintain the secondary maximum. Instead, the secondary maximum is maintained at positive $\mu$ where the community is shaped by competition. It is also interesting to note that there is a reentrant transition as we modify the level of interaction heterogeneity in the community. In (a)-(b), the ecosystem exhibits a secondary maximum and low levels of $\sigma$ and then disappears at moderate $\sigma$. However, the secondary maximum reappears at even higher levels of $\sigma$ with a peak at a higher abundance $N_*$. This region seems to correspond to the RSB region where the ecosystem exhibits multiple steady states.  }
    \label{fig:phase_diagram}
\end{figure}

With the gamut of possible SADs characterized within this stochastic LV model, we now explore the range of macro-ecological parameters that give rise to each class of SADs and the boundaries separating the different regimes. In Fig. \ref{fig:phase_diagram}, we logarithmically sweep through the noise strength $D$ and produce 50 $\times$ 50 phase diagrams in $\mu-\sigma$ space that depicts the location of the secondary maximum, if it exists. Macro-ecological parameters such as the carrying capacity statistics $(K,\sigma_K)$ are fixed at $(1.0,0.2)$, and the migration rate is set such that $\beta\lambda = 0.5$. 

At high levels of stochasticity, predominantly cooperative communities ($\mu <0$) are able to maintain coexistence of many species at high abundances whereas competition results in many species going extinct. Although some mutualism can protect coexistence, it is not without caveats. If mean cooperation is too strong ($\mu < -1$) or if the heterogeneity of interaction strengths $\sigma$ is too high, the quenched statistics of the community, $\overline{\langle N\rangle}$ and $\Delta q = \overline{\langle N^2\rangle - \langle N\rangle^2}$, both diverge. In this region, the ecosystem is unstable and the positive feedback between different species causes all populations to grow unbounded \cite{barbier_cavity_2017,barbier_generic_2018,bunin_ecological_2017,bunin_interaction_2016,biroli_marginally_2018,roy_numerical_2019}. 

However, with decreasing noise strength $D$, a wider range of ecosystems are able to support a local maximum in the SAD. Low levels of mutualism continue to promote the coexistence of many species at very high abundances. As $\mu$ increases towards positive values, the ecosystem becomes more competitive and the local maximum in the species abundance distribution shifts left toward lower abundance. Although competition can give rise to stable coexistence, it is a negative feedback process and a few strong competitors can drive more species toward increasing rarity. Eventually, with stronger competition, the behavior of the SAD shifts from one exhibiting an internal mode to one that resembles the Fisher log-series. At this point, the community consists of many rare or extinct species and very few abundant species. A similar phenomenon occurs as the spread of interaction strength $\sigma$ increases. Even if the mean competition strength $\mu$ is low, a larger spread in the distribution of interaction strengths leads to the presence of more species that can easily outcompete less-fit species. Consequently, there is a transition to a log-series-like SAD as $\sigma$ increases.

However, in the small $D$ limit of predominantly competitive communities ($\mu>0$), increasing the strength and heterogeneity of interactions eventually causes the ecosystem to experience a sharp transition from a single unique equilibrium to a multiple equilibria phase. This is a distinct phase akin to the spin-glass phase in disordered system physics in which the original replica-symmetry assumption breaks down \cite{mezard_spin_1987,nishimori_statistical_2001,biroli_marginally_2018}. However, if the variation in interaction strength $\sigma$ is too great, then the ecosystem can also destabilize with many species being driven to extinction and highly competitive species growing without bound. In this multiple equilibria regime, correlations between species become larger and ever more influential in determining the population size of each species. With the increasing importance of correlations, the community becomes extremely susceptible to small perturbations with minute fluctuations causing drastic changes to the population of each species \cite{mezard_spin_1987,nishimori_statistical_2001,barbier_cavity_2017,barbier_generic_2018,bunin_ecological_2017,bunin_interaction_2016,biroli_marginally_2018}. Details of calculations for the transition between the replica symmetric (RS) and replica symmetry breaking (RSB) phase are found in the Supplemental Information. 

With the assumptions of replica symmetry and negligible correlations breaking down, our calculations of the population statistics in this RSB regime no longer hold. Instead, we must account for the multiple equilibria and the non-Gaussianity of the cavity field distribution $P(h_c)$ to compute the correct abundance statistics and the marginal distribution $P(N)$ \cite{mezard_spin_1987,nishimori_statistical_2001}. Despite the inconsistencies in the replica symmetric calculations, the $P(N)$ in \eqref{eq:marginalDist} still matches the histogram of equilibrium abundances obtained from numerically integrating the Lv equations. This suggests that the RS solution is not drastically different from the RSB one. Hence, we can still use the RS cavity method to obtain a qualitative picture of the behavior of SADs in the RSB regime. In fact, upon closer examination of the theoretical $P(N)$, we find a secondary maximum reemerges in the SAD albeit with a low and broad peak. Since this secondary peak is not very prominent, it becomes difficult to practically distinguish the marginal distribution $P_{S+1}(N)$ from a log-series-like distribution where there is a high representation of species with very low abundances.

 Within highly diverse ecosystems, the residing species often exhibit diverse sets of traits and occupy various niches. These inherent differences between species have long been thought to be the main mechanism shaping the structure of ecosystems, and the statistics should reflect the results from niche theory. Surprisingly, many of the species abundance distributions collected from ecological data match either the Fisher log-series or lognormal-like distribution derived from neutral theory. Although Hubbell's neutral theory imposes an unrealistic assumptions, this agreement between data and neutral theory suggests that processes which are agnostic to species identity such as ecological drift are just as vital in influencing community dynamics.

\floatsetup[figure]{style=plain,subcapbesideposition=top}
\section{Discussion}

To reconcile niche and neutral theory, we constructed a stochastic Lotka-Volterra model of a diverse community that incorporates both niche differentiation and some neutral processes common to many ecosystems. Diversity is captured through drawing each species' carrying capacity and interspecific interactions randomly from a distribution. Although niches are not explicitly modeled, the unique set of interaction strengths for each species already partitions the community into distinct niches. As for neutral processes, we included population noise, environmental noise, and migration from a species pool. 

Using techniques from the physics of disordered systems, we explored the space of macro-ecological parameters and analyzed the population statistics corresponding to each parameter set. The ''phase'' diagram for the stochastic LV model is divided into several distinct regimes, with each characterized by different species abundance distributions: lognormal-like, log-series-like, or multimodal. Local communities with high rates of migration from a larger metacommunity tend to have SADs which have the semblance of a lognormal distribution with many species at intermediate abundances and few species at extremely low and high population sizes. However, once demographic stochasticity gains prominence over migration,
then ecosystem's SAD can behave in two possible ways depending on the characteristics of the interspecific interactions. For a community with uniform and weak competitive interactions, many species die out, but a subset of species survive and coexist at high abundances, resulting in a multimodal SAD.  On the other hand, strong and highly heterogeneous competition generates many strong competitors that drives many species to rarity or even extinction, leaving only a few abundant species. This SAD resembles the Fisher log-series distribution. Interestingly, pushing the mean and variance in competitive strength of the community a little further destabilizes the ecosystem to have multiple equilibrium states and to be sensitive to perturbations. It turns out the SAD transitions back to one with a shallow local peak at intermediate abundances. However, if the heterogeneity in competitive interactions were increased any further, the ecosystem destabilizes again and enters an unbounded growth phase with a few surviving species growing indefinitely. For an ecosystem consisting of predominantly mutualistic interactions between species, many species survive and coexist higher abundances compared to competitive networks of species. This coexistence is remains even at higher levels of noise. However, communities that are strongly cooperative and highly heterogeneous are very unstable, collapsing to an ecosystem with few surviving species growing unbounded.

Using a stochastic Lotka-Volterra niche model, we have shown that neutral statistics need not stem from unrealistic assumptions of species equivalence. Instead they can emerge from the stochastic nature of birth-death processes, migration, and environmental fluctuations. The balance of these neutral processes with the effects of selection generated by niche partitioning and interaction heterogeneity determines whether an ecosystem displays a SAD matching niche theories or neutral theories.

 Interestingly, looking at the $\mu$-$\sigma$ phase diagram at $D=1.0$ in Figure \ref{fig:phase_diagram}(a), the neutral regime with log-series-like has nearly dominated the entire phase diagram while niche phase has nearly disappeared. As the noise strength continues to rise, eventually, the only phase left is the statistically neutral one. Returning to the original un-normalized form of the stochastic LV model in equation \ref{eq:LV_equation}, a noise strength of $\tilde{D} = D/\sqrt{K} \approx 1.0$ corresponds to a model where each species experiences wild population fluctuations of order $\sqrt{KN_i}$. Since the carrying capacity $K$ is quantity generally controlled by the availability of resources and other abiotic factors, then an ecosystem with strong environmental noise would generically yield population statistics and biodiversity patterns that match a neutral model. However, there is an alternative interpretation to the $\sqrt{KN_i}$ fluctuations. Since the abundance of the surviving dominant species $N_i$ tend to be proportional to their carrying capacities $K$, then generic neutral-like statistics is expected to arise when the population dynamics start to have fluctuations of order $\alpha N_i$. This is identical to the noise term of the stochastic Verhulst model, another common logistic model in ecology. In our model where our noise is proportional to $\sqrt{N_i}$, we have implicitly assumed that population fluctuations are predominantly governed by the stochasticity of the linear processes, such as birth in our case. In the stochastic Verhulst model, the $N_i$ scaling commonly arises from the presence of environmental fluctuations, pairwise interactions, quadratic density dependent terms, and other population regulation mechanisms \cite{levins_effect_1969,tuckwell_logistic_1987,hallam_mathematical_1986, kessler_generalized_2015}. With this in mind, we would expect typical diverse ecosystems governed by birth, density dependent death, interactions, and fluctuating environments exhibit phase diagrams where the neutral phase occupies most of the parameter space. This would explain why biodiversity patterns observed across many ecosystems around the world are well-fitted by neutral theory. 
 
  In light of the implicit assumptions in modeling population noise, one fruitful direction would be to study other population noise models with scaling stronger than $\sqrt{N_i}$. In fact, several empirical studies have suggested that noise in some ecosystems scale with the abundance as $N_i^\nu$ where $1/2<\nu<1$. This lies between the pure demographic case and the stochastic Verhulst model \cite{kessler_generalized_2015,kalyuzhny_niche_2014,kalyuzhny_temporal_2014,chisholm_temporal_2014}. Analyzing the $\nu = 1$ case and extracting the corresponding SAD ''phase'' diagram would provide further insight into the prevalence of neutral statistics across ecosystems shaped in part by niche processes.

 Another direction would be to relax the assumptions on the interaction network. Real ecosystems may exhibit particular network structures such as trophic layers present in the food webs \cite{allesina_stabilitycomplexity_2015,allesina_stabilitycomplexity_2015,albert_statistical_2002}. These graphical models require us to establish a degree distribution to describe the number of interactions connected to each node (species), along with a distribution for the interaction strength. Conveniently, the cavity method from disordered systems physics provides an algorithm to solve for the dynamics on networks that are locally tree-like \cite{del_ferraro_cavity_2014,rivoire_cavity_2005,lokhov_dynamic_2015}. Even if the network has interaction loops, the cavity method still provides a good approximation. It would be interesting to explore how network structure, finite connectivity, and distribution of interaction strengths influence the population statistics and the ''phase'' diagram of species abundance distributions.

 \section{Acknowledgments}
 We would like to thank R. Marsland and W. Cui for the valuable and helpful discussions. JW was supported by the National Science Foundation Graduate Research Fellowship Program Grant No. DGE-2039656. DS was supported in part by the US National Science Foundation, through the Center for the Physics of Biological Function (PHY–1734030), by the National Institutes of Health BRAIN initiative (R01EB026943–01); and as a Simons Investigator in MMLS. PM was supported in part as a Simons Investigator in MMLS and NIH NIGMS R35GM119461 grant.

\bibliographystyle{apsrev4-2}
\bibliography{LV_cavity_draft2}

\end{document}


\title{Understanding Species Abundance Distributions in Complex Ecosystems of Interacting Species: Supplementary Material}
\author{Jim Wu, Pankaj Mehta, and David Schwab}

\maketitle

\subsection{Reparameterizing the Stochastic LV Model}
    A diverse ecosystem of $S\gg 1$ interacting species in a fluctuating environment can effectively be modeled by a stochastic Lotka-Volterra (LV) equation of the form
    \begin{align}
        \frac{d\tilde{N}_i}{d\tilde{t}} = \tilde{\lambda} \frac{\tilde{r}_i}{\tilde{K}_i}\tilde{N}_i\left(\tilde{K}_i - \tilde{N}_i - \sum_{j\neq i}^{S} \tilde{\alpha}_{ij}\tilde{N}_j\right) + \sqrt{2\tilde{D}\tilde{N}_i}\eta_i(\tilde{t}) \label{SIeq:LV_equation}
    \end{align}
    Here, each species $i$ grows at a rate $\tilde{r}_i$, immigrates at a rate $\tilde{\lambda}$, has a carrying capacity $\tilde{K}_i = \tilde{K} + \tilde{\sigma}_Kz_K$, and interacts with all other species with strength $\tilde{\alpha}_{ij} = (\tilde{\mu} + \tilde{\sigma} \sqrt{S}z_{ij})/S$. The $z_K$ and $z_{ij}$ are random variables drawn from a standard normal distribution $\mathcal{N}(0,1)$ to capture the heterogeneities and variations between different species. The mean and variance in the carrying capacity are captured by the parameters $\tilde{K}$ and $\tilde{\sigma_K}^2$, while the corresponding statistics of the interaction strengths are summarized by $\tilde{\mu}/S$ and $\tilde{\sigma}^2/S$. The $1/S$ scaling is vital to maintain the stability of the ecosystem. In addition to these deterministic effects, the abundance of each species fluctuate as $\sqrt{2\tilde{D}\tilde{N}_i}$ due to intrinsic birth-death noise and coupling of the population size to environmental noise. In our case, we interpret the multiplicative noise under the Ito convention.
    
    As stated in the main text, equation \eqref{SIeq:LV_equation} can be simplified by making certain reparameterizations and simplifying assumptions about the species parameters that do not change the qualitative picture of the ecosystem. First, since the carrying capacity sets the scale of the population size, we measure the population size in terms of the average carrying capacity $\tilde{K}$. Second, we simplify \eqref{SIeq:LV_equation} by assuming $\tilde{r}_i\tilde{K}/\tilde{K}_i$ to be a constant. This allows us to uniformly rescale the generational time across all species. Subsequently, we reparameterize all the variables:
    \begin{align}
        \tilde{N}_i/\tilde{K} &\longrightarrow N_i\\
        \tilde{t} (\tilde{r}_i\tilde{K}/\tilde{K}_i) &\longrightarrow t\\
        \tilde{K}/\tilde{K} &\longrightarrow 1\\
        \tilde{\sigma}_K/\tilde{K} &\longrightarrow \sigma_K\\
        \tilde{\mu} &\longrightarrow \mu \\
        \tilde{\sigma} &\longrightarrow \sigma\\
        (\tilde{D}/\tilde{K}) &\longrightarrow D\\
        (\tilde{\lambda}/\tilde{K})(\tilde{r}_i\tilde{K}/\tilde{K}_i) &\rightarrow \lambda
    \end{align}
    This effectively reduces \eqref{SIeq:LV_equation} to
    \begin{align}
        \frac{dN_i}{dt} = \lambda +  N_i\left( K_i - N_i-\sum_{j\neq i}^S \alpha_{ij} N_j\right) + \sqrt{2DN_i}\eta_i(t) \label{SIeq:reduced_LV}
    \end{align}

\subsection{Distribution of Species Abundances}
        
    To solve the stochastic LV equation, we can transform \eqref{SIeq:reduced_LV} to the corresponding Fokker-Planck equation (FPE) for the joint probability distribution of all species abundances over time
    \begin{align}
        \frac{\partial}{\partial t} P(\{N_i\};t) &= -\sum_{i=1}^{S} \frac{\partial}{\partial N_i} \left[\lambda + N_i\left( K_i - N_i-\sum_{j\neq i}^S \alpha_{ij} N_j\right) P(\{N_i\};t)\right]  + D\sum^{S}_{i =1}\frac{\partial^2}{\partial N_i^2}\Big( N_i P(\{N_i\};t) \Big) \label{SIeq:FokkerPlanck}
    \end{align}
   We can equivalently write the FPE as a gradient of some probability flux ${\bf J}({\bf N})$
    \begin{align}
       \frac{\partial}{\partial t}P({\bf N};t) &= \nabla_N \cdot {\bf J}({\bf N})  = \nabla_{N}\cdot\Big[- {\bf F}({\bf N}) P({\bf N};t) + D{\bf N}\cdot \nabla_N  P({\bf N};t) + (D-\lambda)P({\bf N};t) \Big]  \label{SIeq:simpleFPE}
    \end{align}
    with ${\bf N} = (N_1,N_2,\ldots, N_i, \ldots, N_S)$ and 
    \begin{align}
        F_i({\bf N}) &= N_i\left( K_i - N_i-\sum_{j\neq i}^S \alpha_{ij} N_j\right)\label{SIeq:F}
    \end{align}
    Since we are interested in the long-time structure of the ecosystem, then we only need the steady-state distribution of the ecosystem. Setting the left-hand side of equation \eqref{SIeq:simpleFPE} to zero, we find that the probability flux ${\bf J}({\bf N})$ must be constant. Recaall that the stationary distribution $P({\bf N})$ is a normalized quantity with assumed finite moments, implying that both $P({\bf N})$ and $\nabla_N P({\bf N})$ must decay to zero as the abundance of any species goes to infinity. Therefore, the probability flux must also be zero when any $N_i\rightarrow \infty$, and since the probability flux is constant, then ${\bf J}({\bf N})$ must be zero everywhere. Hence,
    \begin{align}
        - F_i({\bf N}) P({\bf N}) +DN_i \frac{\partial P({\bf N})}{\partial N_i} + (D-\lambda) P({\bf N})  = 0  \label{SIeq:flux2}
    \end{align}
    and from rearrangement, we have 
   \begin{align}
        \frac{\partial P({\bf N})}{\partial N_i} &= \left[\frac{F_i({\bf N_i})}{D} - \left(1 - \frac{\lambda}{D}\right) \right] \frac{P({\bf N})}{N_i} \label{SIeq:probDisteq}
   \end{align}
    Suppose that the probability distribution has the usual Boltzmann equilibrium form of 
    \begin{align}
        P({\bf N}) = \frac{1}{Z} e^{-\beta H({\bf N})}
    \end{align}
    where $\beta = 1/D$ and $Z$ is the partition function that normalizes the probability distribution. Substituting this into equation \eqref{SIeq:probDisteq}, we obtain
    \begin{align}
        - \beta P({\bf N})\frac{\partial H({\bf N})}{\partial N_i} &= \left(\beta F_i({\bf N_i}) - \left(1 - \beta \lambda\right)\right) \frac{P({\bf N})}{N_i}
    \end{align}
    which simplifies to 
    \begin{align}
        \frac{\partial H({\bf N})}{\partial N_i} &= -\frac{1}{N_i}\left(F_i({\bf N_i}) - \frac{1}{\beta}\left(1 - \beta\lambda\right)\right)
    \end{align}
    Given the form of $F_i({\bf N_i})$ in equation \eqref{SIeq:F}, the Hamiltonian of the system is 
    \begin{align}
        H({\bf N}) &= -\sum_{i=1}^S N_i\left( K_i - \frac{1}{2}N_i - \sum_{j\neq i}^S \alpha_{ij} N_j\right)  + \frac{1-\beta\lambda}{\beta}\ln N_i\label{SIeq:hamiltonian}
    \end{align}
    and thus the joint probability distribution is 
    \begin{align}
        P({\bf N}) = \frac{1}{Z \prod^S_{i=1}N_i^{1-\beta\lambda}}\exp\left[\beta\sum_{i=1}^S N_i\left( K_i - \frac{1}{2}N_i - \sum_{j\neq i}^S \alpha_{ij} N_j\right) \right]\label{SIeq:joint_dist}
    \end{align} 
    
\subsection{The Finite Temperature Cavity Method}
\subsubsection{A Summary of the Cavity Method}
The challenge of extracting the population moments from the joint probability distribution in equation \eqref{SIeq:joint_dist} is due to the heterogeneities in $K_i$ and $\alpha_{ij}$. These are random variables and the goal is to obtain statistics that are independent of the exact values of all $S^2$ parameters. To accomplish this, we must rely on techniques from the disordered systems literature, such as the cavity method. 

The essence of the cavity method is to add (or remove) one species to a large ecosystem already consisting of $S$ species. Every species feels the effect of the new species through the extra interaction term, $h^c = \sum_j \alpha_{0j} N_j$, which disrupts the original population structure and causes it to reshuffle in response. This is called the ``cavity field'' and is denoted by the superscript $c$. Naively under mean-field theory (MFT), the average effect of the new species can be approximated as
\begin{align}
    h_{\text{naive}}^c = \sum^S_{j\neq i} \alpha_{0j} \langle N_j\rangle 
\end{align}
where $\langle \ldots\rangle$ signify the equilibrium average for a given set of $\alpha_{ij}$. The problem with the naive MFT approach is that it misses important fluctuations. If elements of the matrix $\alpha_{ij}$ were equal to $\mu/S$ and scaled as $\mathcal{O}(1/S)$, then the fluctuations in $h^c$ would scale as
\begin{align}
    \delta h^c = \sqrt{\text{Var}(h^c)} = \sqrt{S\left(\frac{\mu}{S}\right)^2\left(\langle N^2\rangle - \langle N\rangle^2\right)} \sim \mathcal{O}(S^{-1/2})
\end{align}
which decays to zero as $S\rightarrow \infty$. However, in our random ecosystem model, the heterogeneities in the pairwise interactions scale as $\mathcal{O}(1/\sqrt{S})$. Hence, the fluctuation in the cavity field is
\begin{align}
    \delta h^c = \sqrt{S\left(\frac{\sigma}{\sqrt{S}}\right)^2\left(\langle N^2\rangle - \langle N\rangle^2\right)} \sim \mathcal{O}(1) \label{SIeq:heterogeneity_scaling}
\end{align}
which is finite in the thermodynamic limit. The simple mean-field approximation fails to account for these $\mathcal{O}(1)$ fluctuations in the cavity field \cite{del_ferraro_cavity_2014,mezard_spin_1987,nishimori_statistical_2001}. 

We can cure mean field theory by making a correction that the cavity field follows a Gaussian distribution. In truth, $h^c$ is not truly Gaussian as the abundances of different species are actually correlated. Increasing the population size of species $i$ generates a reaction in the ecosystem where competing species decrease in abundance and species in mutualistic relationships with species $i$ increase in abundance. However, if these connected correlations $\langle N_iN_j\rangle - \langle N_i\rangle \langle N_j\rangle$ are sufficiently weak, scaling as $\mathcal{O}(1/\sqrt{S})$, then central limit theorem holds and the cavity field $h^c$ is well-approximated by a normal distribution \cite{del_ferraro_cavity_2014,mezard_spin_1987,nishimori_statistical_2001}. 

From this simplification of the interaction term, we can then compute the population statistics of the $S+1$ system.  Taking the thermodynamic limit, the macroscopic observables of the $S+1$ system must be equal to the ones derived for the ecosystem with a species pool of size $S$. From this, we can derive the self-consistent TAP (Thouless-Anderson-Palmer) equations for the population statistics. 

\subsubsection{Computing the Marginal Distribution}

Starting from the joint probability distribution for the abundances of all species in the ecosystem in equation \eqref{SIeq:joint_dist}, we add species $0$ to the ecosystem, yielding the distribution
\begin{align}
    P_{S+1}(\{ N_i\},N_0) \propto P_S(\{N_i\})\times \frac{1}{N_0^{1-\beta\lambda}}\exp\left[\beta N_0 \left(K_0 - \frac{1}{2}N_0 - \sum^S_{j=1} \alpha_{0j}N_j\right)\right] \label{SIeq:joint_add_species}
\end{align}
Notice that the joint distribution for the $S+1$ system is factorized into the original $P(\{N_i\})$ and a Boltzmann factor for the new species. Integrating ove all possible values of the abundances of the original $S$ species, the marginal distribution for species 0 is 
\begin{align}
    P_{S+1}(N_0) \propto \int^\infty_0 \left( \prod_{i=1}^S  dN_i  \right)  P_S(\{N_i\})\times \frac{1}{N_0^{1-\beta\lambda}}\exp\left[\beta N_0 \left(K_0 - \frac{1}{2}N_0 - \sum^S_{j=1} \alpha_{0j}N_j\right)\right] \label{SIeq:marginal}
\end{align}
The only link between species $0$ and the other $S$ species is through the cavity field $h^c = \sum_j \alpha_{0j}N_j$. Hence, we can rewrite the marginal distribution in equation \eqref{SIeq:marginal} as a joint distribution between $N_0$ and $h^c$,
\begin{align}
    P_{S+1}(N_0,h^c) = \frac{1}{\mathcal{N}} \frac{1}{N_0^{1-\beta\lambda}}\exp\left[\beta N_0 \left(K_0 - \frac{1}{2}N_0 - h^c\right)\right] P_S(h^c) \label{SIeq:joint_species_cavity}
\end{align}
where $\mathcal{N}$ is the normalization, and 
\begin{align}
    P_S(h^c) = \int^\infty_{0}\left( \prod^S_{i=1} dN_i\right) \delta\left(h^c - \sum_j \alpha_{ij} N_j\right) P(\{N_i\})
\end{align}
Assuming that the connected correlations between species is negligible, then we can invoke the central limit theorem and propose that the cavity distribution is 
\begin{align}
    P_S(h^c) \approx \frac{1}{\sqrt{2\pi V}} \exp\left(- \frac{(h^c-h_S)^2}{2V}\right)
\end{align}
where the mean and variance of the cavity field are 
\begin{gather}
    \langle h^c\rangle_S = \sum_j \alpha_{0j}\langle N_j\rangle_S \xrightarrow{S\rightarrow\infty} h \label{SIeq:mean_cavity}\\
    \langle (h^c)^2\rangle_S - \langle h^c\rangle^2_S = \sum^S_{j=1}\sum^S_{i=1} \alpha_{0j}\alpha_{0i} \left(\langle N_iN_j\rangle_S - \langle N_i\rangle_S\langle N_j\rangle_S \right) \approx \sum_{i=1}^S \alpha_{0i}^2 \left( \langle N_i^2\rangle - \langle N_i\rangle^2\right) \xrightarrow{S\rightarrow \infty} V \label{SIeq:var_cavity}
\end{gather}
Substituting this back into the joint distribution of $N_0$ and $h^c$, the new marginal distribution for the abundance of species 0 is 
\begin{align}
    P_{S+1}(N_0) &= \frac{1}{\mathcal{N}} \int^\infty_{-\infty}\frac{dh^c}{\sqrt{2\pi V}} \exp\left(- \frac{(h^c-h_S)^2}{2V}\right)\times\frac{1}{N_0^{1-\beta\lambda}} \exp\left[\beta N_0 \left(K_0 - \frac{1}{2}N_0 - h^c\right)\right]\\
        &= \frac{1}{\mathcal{N}}\frac{1}{N_0^{1-\beta\lambda}}\exp\left[\beta N_0 \left(K_0 - \frac{1}{2}N_0 \right)  \right]\exp\left[- \beta N_0 h_S + \frac{1}{2} \beta^2 V N_0^2\right]\\
        &= \frac{1}{\mathcal{N}}\frac{1}{N_0^{1-\beta\lambda}} \exp \left\{- \frac{\beta}{2}\left[ \left(1 - \beta V\right)  N_0^2 - 2\left(K_0 - h_S\right)N_0 \right] \right\} \label{SIeq:marginal2}
\end{align}
The normalization of the marginal distribution $P_{S+1}(N_0)$ is given by
\begin{align}
    \mathcal{N} &= \int^\infty_{0} \frac{dN_0}{N_0^{1-\beta\lambda}} \exp \left\{- \frac{\beta}{2}\left[ \left(1 - \beta V\right)  N_0^2 - 2\left(K_0 - h_S\right)N_0 \right] \right\}\\
        &= \begin{cases}
        (2\beta a)^{-c} \Gamma(2c) U\left(c,\frac{1}{2}, \frac{\beta b^2}{2a}\right) & \text{if } b < 0\\
        \frac{1}{2}\left(\frac{2}{\beta a}\right)^{c}\left[\Gamma\left(c\right) M\left(c,\frac{1}{2}, \frac{\beta b^2}{2a}\right) + 2 \sqrt{\frac{\beta b^2}{2a}} \Gamma\left( \frac{1}{2}+c\right)M\left(\frac{1}{2}+c,\frac{3}{2}, \frac{\beta b^2}{2a}\right) \right] & \text{if } b\geq 0
        \end{cases}
        \label{SIeq:norm_marginal}
\end{align}
where 
\begin{align}
    a &= 1-\beta V\\
    b &= K_0 - h_S\\
    c &= \frac{\beta\lambda}{2}
\end{align}
and $M(x,y,z)$ and $U(x,y,z)$ are the Kummer and Tricomi's confluent hypergeometric functions, respectively. Note that we require $a>0$ and $c>0$ for the integral in \eqref{SIeq:marginal2} to converge.

\subsubsection{Computing the Population Statistics: TAP Equations}
Now that we have the marginal distribution of a representative species in equation \eqref{SIeq:marginal2}, we can extract the population moments
\begin{align}
\langle N_0\rangle_{S+1} &= \frac{1}{\mathcal{N}} \int^\infty_{0} dN_0 ~N_0^{\beta\lambda}\exp \left\{- \frac{\beta}{2}\left[ \left(1 - \beta V\right)  N_0^2 - 2\left(K_0 - h_S\right)N_0 \right] \right\}\label{SIeq:mean}\\
\langle N_0^2\rangle_{S+1} &= \frac{1}{\mathcal{N}} \int^\infty_{0} dN_0 ~N_0^{\beta\lambda + 1}\exp \left\{- \frac{\beta}{2}\left[ \left(1 - \beta V\right)  N_0^2 - 2\left(K_0 - h_S\right)N_0 \right] \right\}\label{SIeq:secondmoment}
\end{align}
Notice that the left-hand side consist of averages over the $S+1$ system while the right-hand side has averages over the $S$ system. Our goal is to obtain self-consistent equations where the averages are only over one system size. We accomplish this by finding relations for the two unknowns left, namely the mean $h$ and variance $V$ of the cavity field. We can determine the mean of the cavity field of the $S+1$ system by multiplying the joint distribution of $N_0$ and $h^c$ in equation \eqref{SIeq:joint_species_cavity} by $h^c$ and integrating over all possible values of the cavity field and the abundance of species 0:
\begin{align}
    h_{S+1} &= \int^\infty_0 dN_0 \int^\infty_{-\infty} dh^c ~h^c P_{S+1}(N_0,h^c) \\
    &= \frac{1}{\mathcal{N}}\int^\infty_0 dN_0 \int^\infty_{-\infty} dh^c \frac{h^c}{N_0^{1-\beta\lambda}} \exp\left[\beta N_0 \left(K_0 - \frac{1}{2}N_0 - h^c\right)\right] \times \frac{1}{\sqrt{2\pi V}} \exp \left[- \frac{(h^c-h_S)^2}{2V} \right]\\
    &= \int^\infty_0 dN_0 \left( h - \beta V N_0 \right) \times  \frac{1}{\mathcal{N}}\frac{1}{N_0^{1-\beta\lambda}} \exp \left\{- \frac{\beta}{2}\left[ \left(1 - \beta V\right)  N_0^2 - 2\left(K_0 - h_S\right)N_0 \right] \right\}\\
    &= h_S - \beta V\langle N_0\rangle_{S+1} \label{SIeq:mean_cavity_field}
\end{align}
The variance in the cavity field $h^c$ as according to equation \eqref{SIeq:var_cavity} is 
\begin{align}
    V =  \sum^S_{i=1}\alpha_{0i}^2 \left(\langle N_i^2\rangle_S - \langle N_i\rangle^2_S\right) 
\end{align}
Taking the quenched average, i.e. averaging over the distribution of $\alpha_{ij}$, we have 
\begin{align}
    \overline{V} &= \sum^S_{i=1}\overline{\alpha_{0i}^2}~ \overline{\left(\langle N_i^2\rangle_S - \langle N_i\rangle^2_S\right) } \\
        &=  \sum^S_{i=1} \overline{\left( \frac{\mu}{S} + \frac{\sigma}{\sqrt{S}} z_{0i}\right)^2} \overline{\left(\langle N_i^2\rangle - \langle N_i\rangle^2\right) } \\
        &= \sum^S_{i=1}\left( \frac{\mu^2}{S^2} + \frac{2\mu\sigma}{S^{3/2}}\overline{z_{0i}} + \frac{\sigma^2}{S}\overline{z_{0i}^2}\right)  \overline{\left(\langle N_i^2\rangle - \langle N_i\rangle^2\right) } \\
        &= \sigma^2\left( \frac{1}{S} \sum^S_{i=1} \overline{ \left(\langle N_i^2\rangle - \langle N_i\rangle^2\right) }\right)  = \sigma^2\Delta q \label{SIeq:var}
\end{align}
In computing the quenched averages of the population fluctuation, we dropped the $S$ subscript in the averages. The difference between the population moments averages over the $S$ and $S+1$ systems is negligible and this is especially true as we take the $S\rightarrow \infty$ limit.

In summary, the TAP equations are 
\begin{align}
    \langle N_0\rangle_{S+1} &= \frac{1}{\mathcal{N}} \int^\infty_{0} dN_0 ~N_0^{\beta\lambda}\exp \left\{- \frac{\beta}{2}\left[ \left(1 - \sigma^2\beta \Delta q\right)  N_0^2 - 2\left(K_0 - \sigma^2\beta \Delta q - h_{S+1}\right)N_0 \right] \right\}\label{SIeq:mean_TAP}\\
\langle N_0^2\rangle_{S+1} &= \frac{1}{\mathcal{N}} \int^\infty_{0} dN_0 ~N_0^{\beta\lambda+1}\exp \left\{- \frac{\beta}{2}\left[ \left(1 - \sigma^2\beta \Delta q\right)  N_0^2 - 2\left(K_0 - \sigma^2\beta\Delta q - h_{S+1}\right)N_0 \right] \right\}\label{SIeq:secondmoment_TAP}\\
\mathcal{N} &= \int^\infty_{0} \frac{dN_0}{N_0^{1-\beta\lambda}}\exp \left\{- \frac{\beta}{2}\left[ \left(1 - \sigma^2\beta \Delta q\right)  N_0^2 - 2\left(K_0 - \sigma^2\beta\Delta q - h_{S+1}\right)N_0 \right] \right\}\label{SIeq:norm_TAP}\\
h_{S+1} &= \sum_{j}\alpha_{0j}\langle N_j\rangle_{S+1} 
\end{align}

\subsubsection{Computing the Population Statistics: Self-Consistent Integral Equations}
Alternatively, we can obtain self-cosnsitent integral equations for all the population moments. Returning to the moments in equations \eqref{SIeq:mean} and \eqref{SIeq:secondmoment}, we know that both $K_0$ and $h_S$ are random with the latter depending on the distribution of $\alpha_{ij}$. So, let us average over the all possible ecosystems, or equivalently, taking the ensemble average over all values of the carrying capacity and the interaction matrix elements. Note that we must take the ensemble average \emph{after} the average over the species abundances. The reasons is that we want to compute the population moments of an ecosystem \emph{given} a particular instance of the carrying capacities $K_i$ and interaction matrix $\alpha_{ij}$, and then obtain the overall statistics independent of the exact values of $K_i$ and $\alpha_{ij}$ via an ensemble average. This is the meaning of a \emph{quenched} average. If we swap the two averages, then we allow the $K_i$ and $\alpha_{ij}$ to equilibrate on the same time scale as the abundances, yielding the \emph{annealed} average instead. 

Now let us compute the quenched average population statistics. Recall that the carrying capacity follows a normal distribution
\begin{align}
    P(K_0) = \frac{1}{\sqrt{2\pi \sigma_K^2}} \exp\left[- \frac{(K_0 -K)^2}{2\sigma_K^2} \right]
\end{align}
with $K = 1$ because we normalized the population size with the average carrying capacity. As for the cavity field, the quenched average of the mean $h_S$ is 
\begin{align}
    \overline{h_S} &= \sum_{j=1}^S \overline{\alpha_{0j}}~\overline{\langle N_j\rangle } = \sum_{j=1}^S \frac{\mu}{S} \overline{\langle N_j\rangle} = \mu \overline{\langle N_j\rangle }
\end{align}
The quenched average of the second moment in $h_S$ is given by 
\begin{align}
    \overline{h^2_S} = \sum^S_{i,j=1}\overline{ \alpha_{0i}\alpha_{0j}} ~\overline{\langle N_i\rangle \langle N_j\rangle } = \sigma^2 \left(\frac{1}{S} \sum_i \langle N_i\rangle^2\right) = \sigma^2 q
\end{align}
since the $\alpha_{0i}$ and $\alpha_{0j}$ are uncorrelated. By the central limit theorem, the cavity field $h_S$ follows a Gaussian:
\begin{align}
    P(h_S) &= \frac{1}{\sqrt{2\pi \sigma^2 \Delta q}} \exp\left[ - \frac{(h_S - \mu \overline{\langle N_j\rangle})^2}{2 \sigma^2 q}\right]
\end{align}
Now that we have the distributions for $K_0$ and $h_S$, let us compute the quenched population moments. Notice in equations \eqref{SIeq:marginal2}, \eqref{SIeq:mean}, and \eqref{SIeq:secondmoment} that we have a difference of two random variables, $u =  K_0 - h_S$. Since both follow normal distributions, then $u$ must also follow a normal distribution as well:
\begin{align} 
    P(u) = \frac{1}{\sqrt{2\pi (\sigma_K^2 + q\sigma^2)}} \exp\left[- \frac{(u - (K-\mu\overline{\langle N\rangle})^2}{2(\sigma_K^2 + q\sigma^2)} \right]
\end{align}
Therefore, the quenched average population size for a representative species is
\begin{align}
    \overline{\langle N_0\rangle} &= \int^\infty_{-\infty} \frac{du}{\sqrt{2\pi (\sigma_K^2+q\sigma^2})}\exp\left[- \frac{(u - (K-\mu\overline{\langle N\rangle})^2}{2(\sigma_K^2 + q\sigma^2)} \right] \times \int^\infty_{0} \frac{dN_0}{\mathcal{N}} ~N_0^{\beta\lambda}\exp \left\{- \frac{\beta}{2}\left[ \left(1 - \sigma^2\beta\Delta q\right)  N_0^2 - 2uN_0 \right] \right\}\\
    &= \int^\infty_{-\infty} Dz \int^\infty_{0} \frac{dN_0}{\mathcal{N}(z)} ~N_0^{\beta\lambda} \exp \left\{- \frac{\beta}{2}\left[ \left(1 - \sigma^2\beta\Delta q\right)  N_0^2 - 2\left(K - \mu \overline{\langle N_i\rangle} +  \sqrt{\sigma_K^2 + q\sigma^2}z\right)N_0 \right] \right\} \label{SIeq:quench_mean}
\end{align}
where we made a change of variables to $z = [u - (K- \mu\overline{\langle N\rangle})]/\sqrt{\sigma_K^2 + q\sigma^2}$ and defined $Dz =dz ~\exp(-z^2/2)/\sqrt{2\pi} $. The normalization $\mathcal{N}(z)$ is given by \eqref{SIeq:norm_marginal} with the same substitution for $u = K_0 - h_S = K-\mu \overline{\langle N\rangle} + \sqrt{\sigma_K^2 + q \sigma^2}z$:
\begin{align}
    \mathcal{N}(z) &= \begin{cases}
        (2\beta a)^{-c} \Gamma(2c) U\left(c,\frac{1}{2}, \frac{\beta b^2}{2a}\right) & \text{if } b < 0\\
        \frac{1}{2}\left(\frac{2}{\beta a}\right)^{c}\left[\Gamma\left(c\right) M\left(c,\frac{1}{2}, \frac{\beta b^2}{2a}\right) + 2 \sqrt{\frac{\beta b^2}{2a}} \Gamma\left( \frac{1}{2}+c\right)M\left(\frac{1}{2}+c,\frac{3}{2}, \frac{\beta b^2}{2a}\right) \right] & \text{if } b\geq 0
        \end{cases}\label{SIeq:quench_norm}
\end{align}
with 
\begin{align}
    a &= 1- \sigma^2\beta\Delta q\\
    b &= K-\mu\overline{\langle N \rangle} + \sqrt{\sigma_K^2 + q\sigma^2}z\\
    c &= \beta\lambda/2
\end{align}
We can repeat the same steps for the quenched average of the second moment in abundance, yielding
\begin{align}
    \overline{\langle N^2_0\rangle} &=\int^\infty_{-\infty} Dz\int^\infty_{0} \frac{dN_0}{\mathcal{N}(z)} ~N_0^{\beta\lambda + 1} \exp \left\{- \frac{\beta}{2}\left[ \left(1 - \sigma^2\beta\Delta q\right)  N_0^2 - 2\left(K - \mu \overline{\langle N_i\rangle} +  \sqrt{\sigma_K^2 + q\sigma^2}z\right)N_0 \right] \right\} \label{SIeq:quench_secondmoment}
\end{align}
Although we have only computed the quenched average statistics for the new species 0, there is no fundamental difference between any of the species. Each species undergo the same ecological processes and exhibit traits all follow similar distributions. Hence, both equations \eqref{SIeq:quench_mean} and \eqref{SIeq:quench_secondmoment} represent the quenched average first and second moment in population size for any species in the ecosystem.

\subsubsection{Summary of Cavity Calculations}

Although we have computed both the TAP equations and the self-consistent integral equations for the population moments, we will be primarily using the latter. Let us summarize all the results computed in the previous subsection and evaluate some of the integrals.

Let us start by defining
\begin{align}
    A &= \frac{\sigma_K^2 + q\sigma^2}{1- \sigma^2\beta \Delta q} \\
    \Delta &= \frac{K - \mu \overline{\langle N\rangle}}{\sqrt{\sigma_K^2 + q\sigma^2}}
\end{align}
which allows us to rewrite the ubiquitous term
\begin{align}
    \frac{\beta (K-\mu\overline{\langle N\rangle} + \sqrt{\sigma_K^2 + q\sigma^2} z)^2}{4(1-\sigma^2\beta\Delta q)} = \frac{\beta A}{4}(z + \Delta)^2
\end{align}
Furthermore, for convenience, let us also define the function
\begin{align}
    G(x,y) = \begin{cases}
        -x \frac{U\left(1 + x,\frac{3}{2},y^2\right)}{U\left(x,\frac{1}{2},y^2\right)} & \text{if } y < 0\\
        \frac{2\Gamma(1 + x)M\left(1+x,\frac{3}{2},y^2\right) + y^{-1} \Gamma\left(\frac{1}{2}+x\right) M\left(\frac{1}{2}+x,\frac{1}{2},y^2\right)}{\Gamma(x)M\left(x,\frac{1}{2},y^2\right) + 2y \Gamma\left(\frac{1}{2}+x\right)M\left(\frac{1}{2}+x,\frac{3}{2},y^2\right)} & \text{if } y\geq 0
        \end{cases}
\end{align}

Now, if we evaluate all the integrals over $N_0$ in equations \eqref{SIeq:quench_mean} and \eqref{SIeq:quench_secondmoment}, and substitute the normalization in \eqref{SIeq:quench_norm}, we find 
\begin{align}
    \overline{\langle N\rangle } &= \frac{\sqrt{\sigma_K^2 + q\sigma^2}}{(1-\sigma^2\beta \Delta q)} \int^\infty_{-\infty} Dz ~(z+b) ~ G\left(\frac{\beta\lambda}{2},\sqrt{\frac{\beta A}{2}}(z+\Delta)\right) \label{SIeq:result_quenchN} \\
    \overline{\langle N^2\rangle } &=  \int^\infty_{-\infty} Dz~\left[\frac{\sigma_K^2 + q\sigma^2}{(1- \sigma^2\beta\Delta q)^2}(z+b)^2 G\left(\frac{\beta\lambda}{2},\sqrt{\frac{\beta A}{2}}(z+\Delta)\right)  + \frac{\beta\lambda}{\beta(1-\sigma^2\beta\Delta q)}  \right]\label{SIeq:result_quenchSqN} 
\end{align}
In addition to these population moments, we need to find self-consistent equations for the quantities $q = \overline{\langle N\rangle ^2}$ and $\Delta q = \overline{\langle N^2\rangle - \langle N\rangle^2}$:
\begin{align}
    q &= \frac{\sigma_K^2 + q\sigma^2}{(1-\sigma^2\beta \Delta q)^2} \int^\infty_{-\infty} Dz ~(z+b)^2 ~G^2\left(\frac{\beta\lambda}{2},\sqrt{\frac{\beta A}{2}}(z+\Delta)\right) \label{SIeq:result_quenchq} \\
    \beta \Delta q &= \int^\infty_{-\infty} Dz \left[ \frac{\beta(\sigma_K^2 + q\sigma^2)}{(1-\sigma^2 \beta \Delta q)^2}(z+b)^2 G\left(\frac{\beta\lambda}{2},\sqrt{\frac{\beta A}{2}}(z+\Delta)\right)\left( 1- G\left(\frac{\beta\lambda}{2},\sqrt{\frac{\beta A}{2}}(z+\Delta)\right) \right) + \frac{\beta\lambda}{(1-\sigma^2\beta\Delta q)}\right] \label{SIeq:result_quenchDeltaq}
\end{align}

Notice that the integrals over the measure $Dz$ are computed over the range $(-\infty,\infty)$. None of the species go extinct in the community due to the steady supply of species from immigration.

\subsection{Asymptotics of the Population Statistics}

To verify these equations for the population statistics, we want to compute the asymptotics of equations \eqref{SIeq:result_quenchN} -\eqref{SIeq:result_quenchDeltaq} and check that they match the $D = 0$ (or equivalently $\beta \rightarrow \infty$) results in \cite{biroli_marginally_2018}. In the limit $\beta \rightarrow \infty$, the quenched average fluctuations in the abundance $\Delta q = \overline{\langle N^2\rangle - \langle N\rangle^2}$ approaches zero. However, it is possible that the quantity $\beta \Delta q$ remains finite. So, the only term that needs to be treated asymptotically is $G\left(x,y\right)$. For large arguments $y$, 
\begin{align}
    G(x,y) \sim \begin{cases}
    \frac{x}{2y} - \frac{x(1+2x)}{4z^2} + \frac{x(1+2x)(3+4x)}{8y^3} - \frac{x(1+2x)(15+34x+20x^2)}{16y^4}  + \mathcal{O}\left( y^{-5}\right) & \text{if } y < 0\\
    1 + \frac{x-\frac{1}{2}}{y} + \frac{x(3-2x)-1}{2y^2} + \frac{(x-1)(2x-1)(4x-5)}{4y^3} - \frac{(x-1)(2x-1)(20x^2-54x + 37)}{8y^4} + \mathcal{O}\left(y^{-5}\right) & \text{if } y\geq 0 
    \end{cases}
\end{align}
When the noise strength $D\rightarrow 0$, or $\beta\rightarrow \infty$, then the case where $y = \sqrt{\frac{\beta A}{2}}(z+\Delta)<0$ goes to zero since the leading term scales as $y^{-1}$, provided that $\beta\lambda$ is small. So, only the case where $y \geq 0$, or $z > -\Delta$ survives. This makes sense since at zero noise, only the species with a positive effective carrying capacity will survive and all those with a negative effective carrying capacity will go extinct. 

Substituting only the zeroth order asymptotic expansion into equations \eqref{SIeq:result_quenchN} - \eqref{SIeq:result_quenchq}, we arrive at 
\begin{align}
    \overline{\langle N\rangle } &= \frac{\sqrt{\sigma_K^2+q\sigma^2}}{1-\sigma^2\beta\Delta q}\int^\infty_{-\Delta}Dz~ (z + \Delta)\label{SIeq:zero_N}\\
    \overline{\langle N^2\rangle } &= \frac{\sigma_K^2 + q\sigma^2}{(1-\sigma^2\beta\Delta q)^2} \int^\infty_{-\Delta}Dz~(z+\Delta)^2 \label{SIeq:zero_SqN}\\
    q &= \frac{\sigma_K^2 + q\sigma^2}{(1-\sigma^2\beta\Delta q)^2} \int^\infty_{-\Delta}Dz~(z+\Delta)^2\label{SIeq:zero_q}
\end{align}
Notice that $\overline{\langle N^2\rangle}$ and $q = \overline{\langle N\rangle^2}$ match in the $\beta \rightarrow \infty$ limit, which agrees with intuition that there are no population fluctuations when noise is absent. As for $\beta\Delta q$, the factor of $\beta$ in the first term of \eqref{SIeq:result_quenchDeltaq} presents a problem. In the small noise limit, $\beta$ will go to infinity while $G(x,y)$ will approach zero if we only asymptotically expand up to zeroth order. Controlling the first term requires a first order asymptotic expansion of $G(\beta\lambda/2,\beta A (z+\Delta)^2/2)$, yielding 
\begin{align}
    \beta \Delta q = \frac{1}{1-\sigma^2\beta\Delta q} \int^\infty_{-\Delta}Dz \label{SIeq:zero_Deltaq}
\end{align}
All of our results agree with the self-consistent equations derived in \cite{biroli_marginally_2018} in the $D\rightarrow 0$ limit.

\subsection{Bifurcation in Species-Abundance Distributions}
 As discussed in the main text, communities dominated by neutral processes with little immigration tend to exhibit unimodal species abundance distributions with a peak at very low populations and a Fisher-log series decay with population size. Neutral-like communities can also exhibit a log-normal distribution, but this occurs only under high immigration as this allows usually rare species to maintain a higher abundance. However, if the ecosystem is partitioned into niches, then the SAD can exhibit multimodality with a peak at low abundance and one or more peaks at a higher population size. 
 
 Depending on the strength of the noise $D$, immigration $\lambda$, the variance in carrying capacity $\sigma_K$, and the mean $mu$ and variance $\sigma^2$ of the interaction strength, the community can transition between different SADs. For example, communities of competitive species in a highly stochastic environment are expected to exhibit SADs that are more akin to neutral theory. On the other hand, at low levels of noise, more species can coexist at higher abundances, resulting in a multimodal SAD found in niche theory. 
 
 Here, we explore how these various SADs can arise from the Lotka-Volterra Model. We start off with a simple model of a community with independent species and no interactions, and show how a bifurcation can arise between neutral-like and niche-like SADs based on the carrying capacity and strength of stochasticity. However, in more realistic ecosystems with interactions, this transition becomes more complicated and the effective carrying capacity is now also shaped by the heterogeneities in species interactions. 

\subsubsection{Stochastic LV equation - No Interactions}
	To illustrate the transition between niche and neutral species abundance distributions, let us start with a simple ecosystem with many independent species and no interactions. The population of each species is modeled by a simple differential equation
	\begin{align}
	\frac{dN}{dt} &= \lambda + N(K-N) + \sqrt{2D N}\eta(t)
	\end{align}
	with $\langle \eta(t)\rangle = 0$ and $\langle \eta(t)\eta(t')\rangle = \delta(t-t')$. The corresponding Fokker-Planck Equation for this Langevin equation with multiplicative noise is 
	\begin{align}
		\frac{d P(N)}{d t} &= - \frac{\partial}{\partial N}\left(\Big(\lambda + N(K-N)\Big)P(N)\right) + D\frac{\partial^2}{\partial N^2}\Big( NP(N)\Big)
	\end{align}
	which has the solution 
	\begin{align}
	P(N) &= \frac{1}{(\mathcal{N}\sqrt{N})^{2}}\exp\left(2\int dN~\frac{\lambda + N(K-N)}{(\sqrt{2DN})^2}\right)\\	
		&= \frac{1}{\mathcal{N} N^{1-\frac{\lambda}{D}}} \exp \left(\frac{KN}{D} - \frac{N^2}{2D} \right)
	\end{align}
	with $\mathcal{N}$ being the normalization. Note that this is a product of a Gaussian and a power-law decay. So, if immigration is small, $(\lambda/D < 1)$, then depending on the parameters $K$ and $D$, the distribution $P(N)$ should exhibit a secondary local maximum in addition to an integrable singularity at $N = 0$. To find the location of this secondary maximum of $P(N)$ at $N_*$, we just need to satisfy
	\begin{align}
	\left.\frac{dP(N)}{dN}\right|_{N = N_*} = 0 \qquad \text{and} \qquad 	\left.\frac{d^2P(N)}{dN^2}\right|_{N = N_*} < 0
	\end{align}
	
	Taking the derivative of $P(N)$ with respect to $N$, we find
	\begin{align}
	0 = \frac{dP(N)}{dN} &= -\left(1-\frac{\lambda}{D}\right)  \frac{1}{\mathcal{N}N^{2-\frac{\lambda}{D}}}\exp \left(\frac{KN}{D} - \frac{N^2}{2D} \right)  + \frac{1}{\mathcal{N}N^{1-\frac{\lambda}{D}}}\left(\frac{K}{D} - \frac{N}{D}\right) \exp \left(\frac{KN}{D} - \frac{N^2}{2D} \right)\\
		&= \left( -\left(1-\frac{\lambda}{D}\right) \frac{1}{N} + \frac{K}{D} - \frac{N}{D}\right)P(N)
	\end{align}
	Since $P(N)$ is not zero, then the expression in the parentheses must be zero. Solving this yields the locations of the local extrema:
	\begin{align}
	N_{*}^{\pm} = \frac{K}{2} \pm \sqrt{\left(\frac{K}{2}\right)^2 - (D-\lambda)}\label{critical}
	\end{align}
	If we look for where the second derivative is negative, then we find the local maximum is at $N_*^{+}$. Notice that we have a local maximum only if the discriminant in \eqref{critical} is greater than zero, i.e. 
	\begin{align}
		\frac{K^2}{4(D-\lambda)} > 1
	\end{align}
	
\subsubsection{Stochastic LV equation - With Interactions}
	
	We can take a similar approach on ecosystems with many species and with a fully connected network of interspecies interactions. We have already solved the corresponding Fokker-Planck equation to the stochastic LV equation and obtained the marginal distribution for a representative species in the ecosystem. Depending on the macro-ecological parameters of the ecosystem, the species abundance distribution can exhibit either niche-like or neutral-like properties with a critical boundary between the two regimes. 
	
	Our dimensionless stochastic LV equation is 
	\begin{align}
	\frac{dN_i}{dt} &= \lambda + N_i\left(K_i - N_i - \sum_{j\neq i}^S \alpha_{ij}N_j\right)+ \sqrt{2DN_i}\eta_i(t)
	\end{align}
    and the marginal distribution for a representative species is
	\begin{align}
	P(N_0,z) = \frac{1}{\mathcal{N} N_0^{1-\beta\lambda}} \exp\left[- \frac{\beta}{2}\left( (1 - \sigma^2\beta\Delta q) N_0^2 -2(K - \mu \overline{\langle N\rangle} + \sqrt{\sigma_K^2 + q\sigma^2}z)N_0\right) \right]
	\end{align}
	where the normalization is 
	\begin{align}
	\mathcal{N}(z) &= \begin{cases}
	(2\beta a)^{-c} \Gamma(2c) U\left(c,\frac{1}{2}, \frac{\beta b^2}{2a}\right) & \text{if } b < 0\\
	\frac{1}{2}\left(\frac{2}{\beta a}\right)^{c}\left[\Gamma\left(c\right) M\left(c,\frac{1}{2}, \frac{\beta b^2}{2a}\right) + 2 \sqrt{\frac{\beta b^2}{2a}} \Gamma\left( \frac{1}{2}+c\right)M\left(\frac{1}{2}+c,\frac{3}{2}, \frac{\beta b^2}{2a}\right) \right] & \text{if } b\geq 0
	\end{cases}
	\end{align}
	with $U(x,y,z)$ and $M(x,y,z)$ being the Tricomi and Kummer's confluent hypergeometric function, respectively, and with
	\begin{align}
		a &= 1- \sigma^2\beta\Delta q\\
		b &= K-\mu\overline{\langle N \rangle} + \sqrt{\sigma_K^2 + q\sigma^2}z\\
		c &= \beta\lambda/2
	\end{align}
	To fully integrate out the interactions, we must marginalize over $z$ to obtain $P(N_0)$:
	\begin{align}
	P(N_0) &= \int  \frac{Dz }{\mathcal{N}(z) N_0^{1-\beta\lambda}} \exp\left[- \frac{\beta}{2}\left( (1 - \sigma^2\beta\Delta q) N_0^2 -2(K - \mu \overline{\langle N\rangle} + \sqrt{\sigma_K^2 + q\sigma^2}z)N_0\right) \right]
	\end{align}
	
	Again, to find the location of the secondary maximum at higher abundance, we must take a derivative of $P(N_0)$ with respect to $N_0$:
	\begin{align}
	\frac{dP(N_0)}{dN_0} &= 
	\Bigg[- \frac{1- \beta\lambda}{N_0} - \beta \left(1-\sigma^2\beta\Delta q\right)N_0 + \beta(K-\mu\overline{\langle N\rangle}) \notag\\
	&\phantom{=} + \beta\sqrt{\sigma_K^2 + q\sigma^2} \left(\frac{\int^\infty_{-\infty} Dz~ \frac{1}{\mathcal{N}(z)}z e^{\beta\sqrt{\sigma_K^2 + q\sigma^2}N_0 z}}{\int^\infty_{-\infty} Dz~\frac{1}{\mathcal{N}(z)} e^{\beta\sqrt{\sigma_K^2 + q\sigma^2}N_0 z}}\right) \Bigg] P(N_0)\\
	&=  \Bigg[- \frac{1- \beta\lambda}{N_0} - \beta \left(1-\sigma^2\beta\Delta q\right)N_0 + \beta(K-\mu\overline{\langle N\rangle}) \notag\\
	&\phantom{=} + \beta\sqrt{\sigma_K^2 + q\sigma^2} \left(\frac{\int^\infty_{-\infty} ~ \frac{Dy~y}{\mathcal{N}(y+\beta\sqrt{\sigma_K^2 + q\sigma^2}N_0)}}{\int^\infty_{-\infty}~\frac{Dy}{\mathcal{N}(y+\beta\sqrt{\sigma_K^2 + q\sigma^2}N_0)}} +  \beta \sqrt{\sigma_K^2 + q\sigma^2}N_0 \right) \Bigg] P(N_0) \label{SIeq:derivP}
	\end{align}
    Setting the first derivative to zero, we find that a secondary maximum in $P(N)$ exists only if the equation
    \begin{align}
       ( 1- \beta\lambda) + \left(1-\sigma^2\beta\Delta q\right)N_0^2 - \left(K-\mu\overline{\langle N\rangle} + \sqrt{\sigma_K^2 + q\sigma^2} \left(\frac{\int^\infty_{-\infty} Dz~ \frac{1}{\mathcal{N}(z)}z e^{\beta\sqrt{\sigma_K^2 + q\sigma^2}N_0 z}}{\int^\infty_{-\infty} Dz~\frac{1}{\mathcal{N}(z)} e^{\beta\sqrt{\sigma_K^2 + q\sigma^2}N_0 z}}\right) \right)N_0 = 0 \label{SIeq:secondmax}
    \end{align}
    has positive real solutions. It turns out that if the secondary maximum exists, equation \eqref{SIeq:secondmax} has two real solutions, one for the local minimum and another corresponding to the local maximum. The larger solution $N_*$ is always the local maximum and this can be verified by checking if the second derivative of $P(N_0)$ is negative at $N_*$.

\subsection{Iterative Algorithm for Solving the Self-Consistent Equations}

Analytically solving the self-consistent equations for the population statistics in \eqref{SIeq:result_quenchN} - \eqref{SIeq:result_quenchDeltaq} is intractable due to the ratio of sums of modified Bessel functions in $F(x)$. Instead, we can still extract the desired statistics by numerically solving the integral equations via an iterative scheme.

The algorithm starts by initializing the quenched population statistics $A = \{A_i\} =  \{ \overline{\langle N \rangle},\overline{\langle N^2\rangle}, q, \beta \Delta q\}$ with random values in the interval $[0,1]$ since the normalized population size tends to be in this range. Substituting thise initial values $A_{old}$ into the integrals in \eqref{SIeq:result_quenchN} - \eqref{SIeq:result_quenchDeltaq}, we compute new values $A_{new}$. We update the population statistics by taking a linear combination of $A_{old}$ and the $A_{new}$ with a parameter $\alpha$:
\begin{align}
    A_{updated} = (1-\alpha) A_{old} + \alpha A_{new}
\end{align}
By setting $\alpha$ to be in the range $[0,1]$, we dampen any wild oscillations in $A$ over consecutive iterations and allow the solution to settle more gently to their true values. We find that $\alpha = 0.3$ works fairly well \cite{roy_numerical_2019}. 

The iterative scheme stops when the $L^2$ norm of the difference between $A_{old}$ and $A_{new}$ satisfies the convergence criteria
\begin{align}
    L = \sqrt{\sum_{i=1}^4 (A^i_{old} - A^i_{new})^2} < \theta
\end{align}
where $\theta$ is the threshold. In our computations, we set $\theta = 1\times 10^{-7}$, and the population statistics tend to converge quite quickly independent of the initialization of $A$. However, if the convergence criteria is not satisfied within 1000 iterations, then there is no equilibrium solution and the ecosystem is deemed to have collapsed.

Using this algorithm, we swept through a $50\times 50$ grid of $\mu-\sigma$ space for various values of $D$. For $D = 0$, since the integral equations are much easier to numerically compute, we can sweep over a larger $500\times500$ grid instead. Similarly, we sweep over $50\times 50$ grids of $D-\sigma$ space for various values of $\mu$ to obtain a more complete picture of the niche-neutral phase diagrams.

\subsection{Numerical Integration of Stochastic LV Equations}

In addition to the solving the self-consistent integral equations, we also numerically integrate the stochastic Lotka-Volterra equations in \eqref{SIeq:reduced_LV} to verify our results. We created a community of $S = 1000$ species, randomly assigning each species with a carrying capacity from a normal distribution $\mathcal{N}(1,\sigma_K^2)$ and initializing each species with a random abundance from a uniform distribution. We also generate an interaction matrix with each element drawn from a Gaussian distribution $\mathcal{N}(\mu/S,\sigma^2/S)$. 

Most numerical integration schemes for stochastic differential equations (SDE), such as the Euler-Maruyama and Milstein method, are built for the equations interpreted under the Ito convention. The non-anticipating nature of the noise in the Ito convention makes it ideal and easier to implement in numerical computations. Nevertheless, we can still easily modify the Milstein method for Stratonovich SDEs. For a Stratonovich SDE of the form 
\begin{align}
   dX_t  = a(X_t)dt + b(X_t)\circ dW_t
\end{align}
the corresponding Ito SDE is 
\begin{align}
    dX_t = \left(a(X_t) + \frac{1}{2}b(X_t) b'(X_t) \right) + b(X_t)dW_t
\end{align}
where $W_t$ represents a Wiener process with $\Delta W_t = W_{t+1} - W_t$ distributed normally with mean zero and variance $\Delta t$. Hence, the Milstein scheme for this SDE is 
\begin{align}
    X_{t+1} &= X_t + \left( a(X_t) + \frac{1}{2}b(X_t)b'(X_t) \right) \Delta t + b(X_t)\Delta W_t + \frac{1}{2}b(X_t)b'(X_t) \left((\Delta W_t)^2 - \Delta t \right)\\
        &= X_t + a(X_t) \Delta t + b(X_t)\Delta W_t + \frac{1}{2}b(X_t)b'(X_t) (\Delta W_t)^2 
\end{align}
We can generalize this for system of Stratonovich SDEs when we have diagonal noise, i.e. when the Wiener process for each species is independent of each other. If there were off-diagonal terms coupling the dynamics of one variable to the noise of another, then the final term in the Milstein method would be a bit more complicated. For our stochastic LV equation, the Milstein scheme is 
\begin{align}
    N_i^{t+1} = N_i^t + N_i^t\left(K_i - N_i^t - \sum^S_{j\neq i} \alpha_{ij} N_j^t \right)\Delta t + \sqrt{2D N_i^t} \Delta W_t +  \frac{D}{2}(\Delta W_t)^2 
\end{align}
The system of SDEs was integrated using the Milstein method over $T = 10^5$ time steps with a step size of $\Delta t = 0.005$. If the population of any species goes below zero at any point, the abundance is set to zero for all subsequent times. For the surviving species, once their abundances start to plateau, we use the last half of the time points to compute the steady state population moments for each species. These simulations were replicated $R = 30$ times with different interaction matrices and carrying capacity vectors sampled from the same distributions. With these $S\times R = 30,000$ simulated species and their corresponding population moments, we can construct a histogram for the probability density of species abundances. Furthermore, we can compute the quenched moments by averaging the population moments of the 30,000 simulated species. We can then compare the simulated distributions and the quenched population statistics to the ones computed theoretically from the cavity method as shown in Figures \ref{fig:marginal_dist_comp} and \ref{fig:theory_sim_comp}.

\subsection{Replica-Symmetry Breaking (RSB)}
In addition to the niche-neutral transition, there is another transition corresponding to replica-symmetry breaking (RSB) where the ecosystem can exhibit multiple equilibrium solutions. In the replica symmetry scenario, we assumed that the connected correlation between two different species is negligible and only used the variance in abundances to compute the quenched population statistics. However, if we include these $1/\sqrt{S}$ correlations between species abundances, then the replica symmetric solution becomes unstable and gives rise to new solutions. 

To find the multiple equilibria phase, we need to analyze a system of size $S + 2$, denoted by $N_0$ and $N_{0'}$ \cite{del_ferraro_cavity_2014,mezard_spin_1987}. The local fields associated with each of the two species are similarly $h_0$ and $h_{0'}$ and they will be correlated 
\begin{align}
	\langle h_0 h_{0'}\rangle_c = \langle h_0h_{0'}\rangle - \langle h_0\rangle \langle h_{0'} \rangle= \sum_{i,j=1}^{S}\alpha_{0i}\alpha_{0'j}\langle N_i N_j \rangle_c\sim \mathcal{O}(1/\sqrt{S})
\end{align}
Instead of a bivariate normal distribution of independent variables, We assume that the probability distribution is of the form 
\begin{align}
	\mathcal{P}(h_0,h_{0'}) &= \mathcal{N}(\boldsymbol{\mu},\boldsymbol{\Sigma}) = \frac{1}{\sqrt{(2\pi)^2\det\boldsymbol{\Sigma}}} \exp\left(-\frac{1}{2}(\mathbf{h}-\boldsymbol{\mu})^T \boldsymbol{\Sigma}^{-1} (\mathbf{h}-\boldsymbol{\mu})\right) \label{multiGauss}
\end{align}
where 
\begin{align}
	\mathbf{h} = \begin{pmatrix}
	h_0\\
	h_{0'}
	\end{pmatrix} \quad,\quad \boldsymbol{\mu} = \begin{pmatrix}
	\langle h_0\rangle\\
	\langle h_{0'}\rangle 
	\end{pmatrix}
	\quad , \quad
	\boldsymbol{\Sigma}^{-1} = \begin{pmatrix}
		\frac{1}{V} & -\epsilon\\
		-\epsilon & \frac{1}{V}
	\end{pmatrix}
\end{align}
so that the probability distribution can be rewritten in the form 
\begin{align}
\mathcal{P}(h_0,h_{0'}) &\sim \exp\left(\epsilon(h_0 - \langle h_0\rangle)(h_{0'} - \langle h_{0'}\rangle)\right)\exp\left(-\frac{(h_0-\langle h_0\rangle)^2}{2V}\right)\exp\left(-\frac{(h_{0'}-\langle h_{0'}\rangle)^2}{2V}\right)
\end{align}
Notice the products of Gaussians, which would represent the joint probability distributions of the local fields at species $0$ and $0'$ if they were independent from each other. The term in front introduces a small correlation between the two species. Inverting $\boldsymbol{\Sigma}^{-1}$, we find that
\begin{align}
	\boldsymbol{\Sigma} = \frac{1}{(1/V)^2 - \epsilon^2} \begin{pmatrix}
	\frac{1}{V} & \epsilon\\
	\epsilon & \frac{1}{V}
	\end{pmatrix}
\end{align}
and hence the correlations between $h_0$ and $h_{0'}$ is 
\begin{align}
	\langle h_0h_{0'}\rangle_c = \frac{V^2\epsilon}{1-V^2\epsilon^2}\approx V^2\epsilon
\end{align}
provided that $\epsilon$ is small.

Now returning back to the LV model, by adding a new species $0$ and $0'$, the joint probability distribution of $y_0,y_{0'},h_0, h_{0'}$ is 
\begin{align}
	\mathcal{P}(N_0,N_{0'},h_0, h_{0'}) &= \frac{1}{Z\sqrt{N_0N_{0'}}}\exp\left( \beta N_0\left(K_0 - \frac{1}{2}N_0 - h_0\right) \right)\exp\left( \beta N_{0'}\left(K_{0'} - \frac{1}{2}N_{0'}- h_{0'}\right) \right) \exp\left(- \beta\alpha_{00'}N_0N_{0'}\right)\notag \\
		&\phantom{==} \times \exp\left(\epsilon(h_0 - \langle h_0\rangle)(h_{0'} - \langle h_{0'}\rangle)\right)\exp\left(-\frac{(h_0-\langle h_0\rangle)^2}{2V}\right)\exp\left(-\frac{(h_{0'}-\langle h_{0'}\rangle)^2}{2V}\right)
\end{align}
Now we must integrate out $h_0$ and $h_{0'}$:
\begin{align}
	P(N_0,N_0') ~&\propto \int^\infty_{-\infty}\int^\infty_{-\infty} e^{-\beta N_0 h_0 - \beta N_{0'}h_{0'}} \mathcal{P}(h_0,h_{0'})dh_0dh_{0'}\\	
		&=\int^\infty_{-\infty}\int^\infty_{-\infty} e^{-\frac{y_0^2}{D}h_0 - \frac{y_{0'}^2}{D}h_{0'}}  e^{\epsilon(h_0 - \langle h_0\rangle)(h_{0'} - \langle h_{0'}\rangle)}\exp^{-\frac{(h_0-\langle h_0\rangle)^2}{2V}-\frac{(h_{0'}-\langle h_{0'}\rangle)^2}{2V}}dh_0dh_{0'}
\end{align}
This integral is identical to the moment generating function $\langle e^{\mathbf{t}\mathbf{h}}\rangle$ where $\mathbf{t} =- \begin{pmatrix}
y_0^2/D\\ y_{0'}^2/D
\end{pmatrix}$. For a multivariate Gaussian, the moment generating function is 
\begin{align}
	M(\mathbf{t}) &= \exp\left( \boldsymbol{\mu}^T \mathbf{t} + \frac{1}{2}\mathbf{t}^T\boldsymbol{\Sigma}\mathbf{t}\right)
\end{align}
Therefore, the joint distribution between $N_0$ and $N_{0'}$ is 
\begin{align}
	P(N_0,N_{0'}) &\approx  \frac{1}{Z(N_0N_{0'})^{1-\beta\lambda}}\exp\left(\beta N_0\left(K_0-\frac{1}{2}N_0\right)\right)\exp\left(\beta N_{0'}\left(K_{0'}-\frac{1}{2}N_{0'}\right)\right)\exp\left(-\beta \left(N_0\langle h_0\rangle + N_{0'}\langle h_{0'}\rangle\right) \right)\notag\\
	&\phantom{==} \times\exp\left(-\beta\alpha_{00'}N_0N_{0'}\right) \exp\left( \beta^2V^2\left( \frac{N_0^2}{2V} + \frac{N_{0'}^2}{2V} + \epsilon N_0N_{0'} \right)  \right)\\
	& = \frac{1}{Z(N_0N_{0'})^{1-\beta\lambda}}\exp\left[- \frac{\beta}{2}\left(\left(1 - \beta V \right)(N_0^2 +N_{0'}^2) -  2(K_0 -\langle h_0\rangle)N_0 -2 (K_{0'}-\langle h_{0'}\rangle)N_{0'}\right)\right]\notag\\
	&\phantom{==} \times \exp\left[\beta(\beta V^2\epsilon -\alpha_{00'}) N_0N_{0'}   \right]
 \end{align}
 Since $V^2\epsilon - \alpha_{00'}$ is very small, then we can approximate the joint distribution as 
 \begin{align}
	 P(N_0,N_{0'}) &\approx  \frac{1}{Z(N_0N_{0'})^{1-\beta\lambda}}\exp\left[- \frac{\beta}{2}\left(1 - \beta V \right)(N_0^2 +N_{0'}^2) + \beta(K_0 -\langle h_0\rangle)N_0 + \beta (K_{0'}-\langle h_{0'}\rangle)N_{0'} \right] \notag\\
	 &\phantom{==} \times \left(1 + \beta (\beta V^2\epsilon-\alpha_{00'})N_0N_{0'}\right)
 \end{align}
  Hence the effective partition function is 
 \begin{align}
 	Z &\approx \left(\int^\infty_0 \frac{dN_0}{N_0^{1-\beta\lambda}}\exp\left[-\frac{\beta}{2}\left(1-\beta V\right)N_0^2 + \beta(K_0 - \langle h_0\rangle)N_0\right]\right) \left(\int^\infty_0 \frac{dN_{0'}}{N_{0'}^{1-\beta\lambda}}\exp\left[-\frac{\beta}{2}\left(1-\beta V\right)N_{0'}^2 + \beta(K_{0'} - \langle h_{0'}\rangle)N_{0'}\right]\right) \notag\\
 	&\phantom{==} + \beta(\beta V^2\epsilon-\alpha_{00'})\left(\int^\infty_0 dN_0 ~N_0^{\beta\lambda}\exp\left[-\frac{\beta}{2}\left(1-\beta V\right)N_0^2 + \beta(K_0 - \langle h_0\rangle)N_0\right]\right)\notag\\
 	&\phantom{====}\times \left(\int^\infty_0 dN_{0'} ~N_{0'}^{\beta\lambda}\exp\left[-\frac{\beta}{2}\left(1-\beta V\right)N_{0'}^2 + \beta(K_{0'} - \langle h_{0'}\rangle)N_{0'}\right]\right)\\
 	&= Z_0^{RS}Z_{0'}^{RS}\left( 1+ \beta(\beta V^2\epsilon - \alpha_{00'})\langle N_0\rangle^{RS}\langle N_{0'}\rangle ^{RS}\right)
 \end{align}
 with the $RS$ denoting the replica-symmetric version. This means that we can rewrite the joint probability distribution as 
 \begin{align}
     P(N_0,N_{0'}) &= \frac{\left(1 + \beta \left(\beta V^2\epsilon- \alpha_{00'}\right)N_0N_{0'}\right) }{\left( 1+ \beta(\beta V^2\epsilon - \alpha_{00'})\langle N_0\rangle_{RS}\langle N_{0'}\rangle_{RS}\right)} P_{RS}(N_0)P_{RS}(N_{0'})\\
    &\approx \left( 1 -  \beta(\beta V^2\epsilon - \alpha_{00'})\langle N_0\rangle_{RS}\langle N_{0'}\rangle_{RS} + \beta\left(\beta V^2\epsilon- \alpha_{00'}\right) N_0N_{0'}\right)P_{RS}(N_0)P_{RS}(N_{0'})
 \end{align}
Since $\beta (\beta V^2\epsilon -\alpha_{00'})$ is small. From this, we can compute the connected correlation:
\begin{align}
 \langle N_0N_{0'}\rangle^c_{RSB}&= \langle N_0 N_{0'}\rangle_{RSB} - \langle N_0 \rangle_{RSB} \langle N_{0'}\rangle_{RSB} \\
 	&= \int^\infty_0 dN_0 dN_{0'} ~N_0 N_{0'}P(N_0,N_{0'})  - \left(\int^\infty_0 dN_0 dN_{0'} ~N_0 P(N_0,N_{0'})\right) \left(\int^\infty_0 dN_0 dN_{0'} ~ N_{0'}P(N_0,N_{0'})\right) \\
 		&=\Big(\langle N_0\rangle_{RS} \langle N_{0'}\rangle_{RS} - \beta(\beta V^2\epsilon-\alpha_{00'})\langle N_0\rangle_{RS}\langle N_{0'}\rangle_{RS}+ \beta(\beta V^2\epsilon  - \alpha_{00'})\langle N_0^2\rangle_{RS} \langle N_{0'}^2\rangle_{RS} \Big) \notag\\
 		&\phantom{=} - \Big(\langle N_0\rangle_{RS} - \beta(\beta V^2\epsilon  - \alpha_{00'})\langle N_0\rangle_{RS}  + \beta(\beta V^2\epsilon  - \alpha_{00'})\langle N_{0'}\rangle_{RS}\langle N_0^2\rangle_{RS}  \Big)\notag\\
 		&\phantom{===}\times \Big(\langle N_{0'}\rangle_{RS}- \beta(\beta V^2\epsilon  - \alpha_{00'}) \langle N_{0'}\rangle_{RS} + \beta(\beta V^2\epsilon  - \alpha_{00'})\langle N_0\rangle_{RS}\langle N_{0'}^2\rangle_{RS}\Big) \\
 		&\approx \beta(\beta V^2\epsilon  - \alpha_{00'})\Big(\langle N_0\rangle^2_{RS}\langle N_{0'}\rangle^2_{RS} - \langle N_0\rangle^2_{RS}\langle N_{0'}^2 \rangle_{RS} - \langle N_0^2\rangle_{RS} \langle N_{0'}\rangle^2_{RS} + \langle N_0^2\rangle_{RS} \langle N_{0'}^2\rangle_{RS} \Big)\\
 		&=  \beta(\beta V^2\epsilon  - \alpha_{00'}) \Big( \langle N_0^2\rangle_{RS} -\langle N_0\rangle^2_{RS}\Big)\Big(\langle N_{0'}^2\rangle_{RS} - \langle N_{0'}\rangle^2_{RS}\Big) 
 \end{align}
 Substituting our definition for the correlation between the local fields for species $0$ and $0'$,
 \begin{align}
 	\langle N_0N_{0'}\rangle^c &= \left(\beta(\beta \langle h_0h_{0'}\rangle_c - \alpha_{00'})\right) \left( \langle N_0^2\rangle_{RS}-\langle N_0\rangle^2_{RS}\right)\left(\langle N_{0'}^2\rangle_{RS} - \langle N_{0'}\rangle^2_{RS}\right) \\
 		&=  \beta \left(\beta\sum_{jk}\alpha_{0j}\alpha_{0'k}\langle N_j N_{k}\rangle^c - \alpha_{00'}\right) \left( \langle N_0^2\rangle_{RS} -\langle N_0\rangle^2_{RS}\right)\left(\langle N_{0'}^2\rangle_{RS} - \langle N_{0'}\rangle^2_{RS}\right)  \label{SIeq:corrFunc}
 \end{align}
Recalling that the term in the parenthesis of equation \eqref{SIeq:corrFunc} is small, then sample average of the correlation vanishes, i.e. $\langle N_0N_{0'}\rangle^c \rightarrow 0$ as $M\rightarrow \infty$ where $M$ is the number of samples or replicas of the matrix $\alpha_{ij}$ drawn. Hence, we need to compute the sample average of the correlation squared, yielding the nonlinear spin-glass correlation function. Just squaring equation \eqref{SIeq:corrFunc} yields 
 \begin{align}
 	(\langle N_0N_{0'}\rangle^c)^2&= \beta^2\left(\alpha_{00'}^2 + \beta^2 \sum_{jklm}\alpha_{0j}\alpha_{0'k}\alpha_{0l}\alpha_{0'm}\langle N_j N_k\rangle^c \langle N_lN_m\rangle^c  -2\alpha_{00'}\sum_{jk}\alpha_{0j}\alpha_{0'k}\langle N_jN_k\rangle_c\right)\\
 		&\phantom{==}\times \left( \langle N_0^2\rangle_{RS} -\langle N_0\rangle^2_{RS}\right)^2\left(\langle N_{0'}^2\rangle_{RS} - \langle N_{0'}\rangle^2_{RS}\right) ^2
 \end{align}
 and taking the ensemble average yields
 \begin{align}
 	\overline{ (\langle N_0N_{0'}\rangle^c)^2} &= \beta^2\left(\overline{ \alpha_{00'}^2}+ \beta^2 \sum_{jk} \overline{\alpha_{0j}^2}\overline{\alpha_{0'k}^2}\overline{ (\langle N_jN_k\rangle^c)^2} + \beta^2 \sum_{jl} \overline{\alpha_{0j}\alpha_{0l}} \cdot \overline{\alpha_{0'j} \alpha_{0'l}} \overline{\langle N_j^2\rangle^c \langle N_l^2\rangle^c}\right) \notag\\
 		&\phantom{==}\times \overline{\left( \langle N_0^2\rangle_{RS} -\langle N_0\rangle^2_{RS}\right)^2\left(\langle N_{0'}^2\rangle_{RS} - \langle N_{0'}\rangle^2_{RS}\right)^2 }\\
 		&= \beta^2\left(\frac{\sigma^2}{S} + \beta^2\sigma^4\overline{ \langle N_jN_k\rangle^2}_c + \frac{\beta^2\sigma^4}{S}\overline{ (\langle N_{j}^2\rangle^c)^2}\right) \overline{\left( \langle N_0^2\rangle^c_{RS} \right)^2\left(\langle N_{0'}^2\rangle^c_{RS}\right)^2 }\\
 		&= \beta^2\left(\frac{\sigma^2}{S} + \beta^2\sigma^4 \overline{ (\langle N_0N_{0'}\rangle^c)^2} + \frac{\beta^2\sigma^4}{S}\overline{ (\langle N_0^2 \rangle^c)^2}\right)\overline{\left( \langle N_0^2\rangle^c_{RS} \right)^2}^2
 \end{align}
 In the last line, we work in the thermodynamic limit, which means that the $S+2$ and the $S$ ecosystem must yield equivalent macroscopic quantities, i.e. $\overline{ \langle N_0N_{0'}\rangle^2_c} = \overline{ \langle N_jN_{k}\rangle^2_c}$ and $\overline{\left( \langle N_0^2\rangle^c_{RS} \right)^2\left(\langle N_{0'}^2\rangle^c_{RS}\right)^2 } = \overline{\left( \langle N_0^2\rangle^c_{RS} \right)^2}^2$. Now, solving for desired square of the connected correlation and dropping the $RS$ subscript, we find 
 \begin{align}
 	\overline{(\langle N_0N_{0'}\rangle^c)^2} &= \beta^2 \frac{\left(\frac{\sigma^2}{S}\right)\left(1 + \beta^2 \sigma^2\overline{ \langle N_0^2 \rangle_c^2}\right)}{1 - \beta^4\sigma^4 \left(\overline{ \langle N_0^2 \rangle_c^2}\right)^2}\overline{ \langle N_0^2 \rangle_c^2} = \frac{\beta^2\sigma^2/S}{1-\beta^2\sigma^2\left[\overline{ \left(\langle N_0^2 \rangle -\langle N_0\rangle^2\right)^2}\right]} \left[\overline{ \left(\langle N_0^2 \rangle -\langle N_0\rangle^2\right)^2}\right]^2
 \end{align}
 Hence, in the $S \rightarrow \infty$ limit, the nonlinear spin-glass susceptibility is
\begin{align}
	\chi_2 = \frac{1}{\beta^2S} \sum_{kl}\langle N_kN_l\rangle_c^2 \rightarrow  \frac{\sigma^2}{1-\beta^2\sigma^2\left[\overline{ \left(\langle N_0^2 \rangle -\langle N_0\rangle^2\right)^2}\right]} \left[\overline{ \left(\langle N_0^2 \rangle -\langle N_0\rangle^2\right)^2}\right]^2
\end{align}
 For the nonlinear susceptibility to remain positive definite and finite, we require the stability condition
 \begin{align}
 	x = 1- \beta^2\sigma^2 \left[\overline{ \left(\langle N_0^2 \rangle -\langle N_0\rangle^2\right)^2}\right] > 0
 \end{align}
 Substituting the equations for $\langle N_0^2\rangle$ and $\langle N_0\rangle$ from the cavity method, we find that the replica symmetric solution is stable only when 
 \begin{align}
  1 -  \sigma^2\int^\infty_{-\Delta} Dz \left[ \frac{\beta(\sigma^2_K + q\sigma^2)}{(1-\sigma^2\beta\Delta q)^2}(z+\Delta)^2G\left(\frac{\beta\lambda}{2},\frac{\beta A}{2}(z+\Delta)^2\right)\left(1-G\left(\frac{\beta\lambda}{2},\frac{\beta A}{2}(z+\Delta)^2\right) \right) + \frac{\beta\lambda}{1-\sigma^2\beta\Delta q} \right]^2 \geq 0 \label{stability}
 \end{align}
 
In the low temperature limit, $\beta\rightarrow\infty$, and when the immigration rate is low,  this stability criteria reduces to

9 \begin{align}
 	1 - \frac{\sigma^2}{(1-\beta\sigma^2\Delta q)^2} \int_{-\Delta}^\infty Dz  \geq 0 \label{SIeq:RSB}
 \end{align}
which matches the replica-symmetry breaking results in \cite{biroli_marginally_2018}. 

\begin{figure}
    \centering
    \sidesubfloat[]{\includegraphics[width=0.7\textwidth]{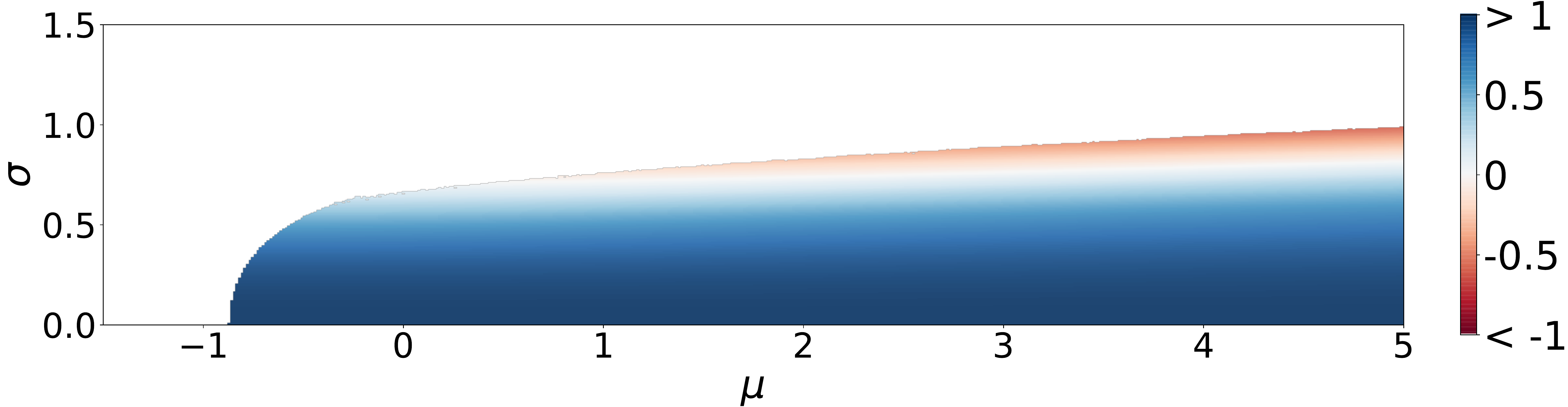}}\\
    \sidesubfloat[]{\includegraphics[width=0.7\textwidth]{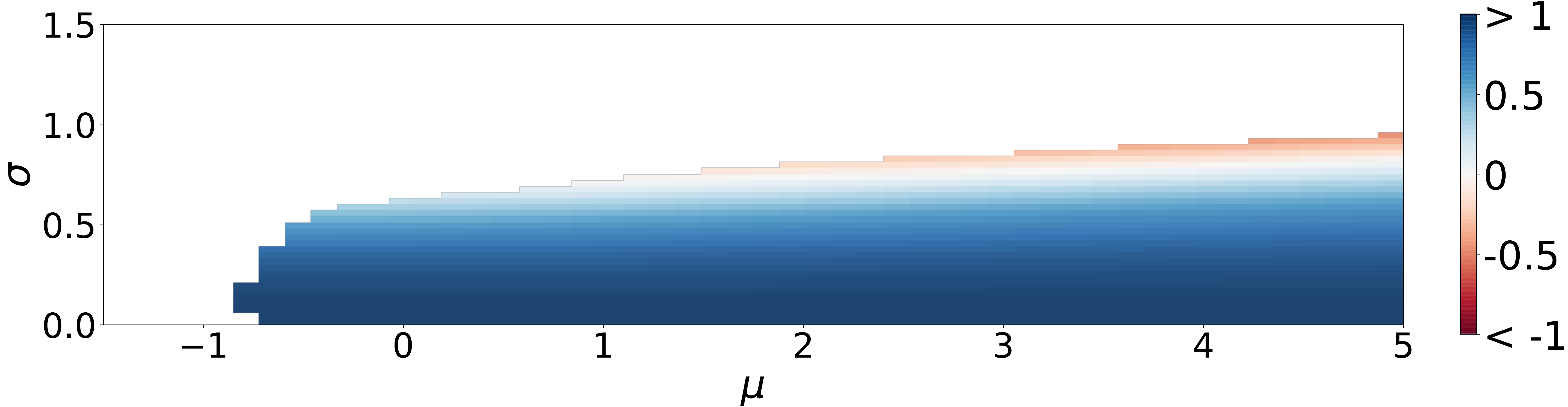}}\\
    \sidesubfloat[]{\includegraphics[width=0.7\textwidth]{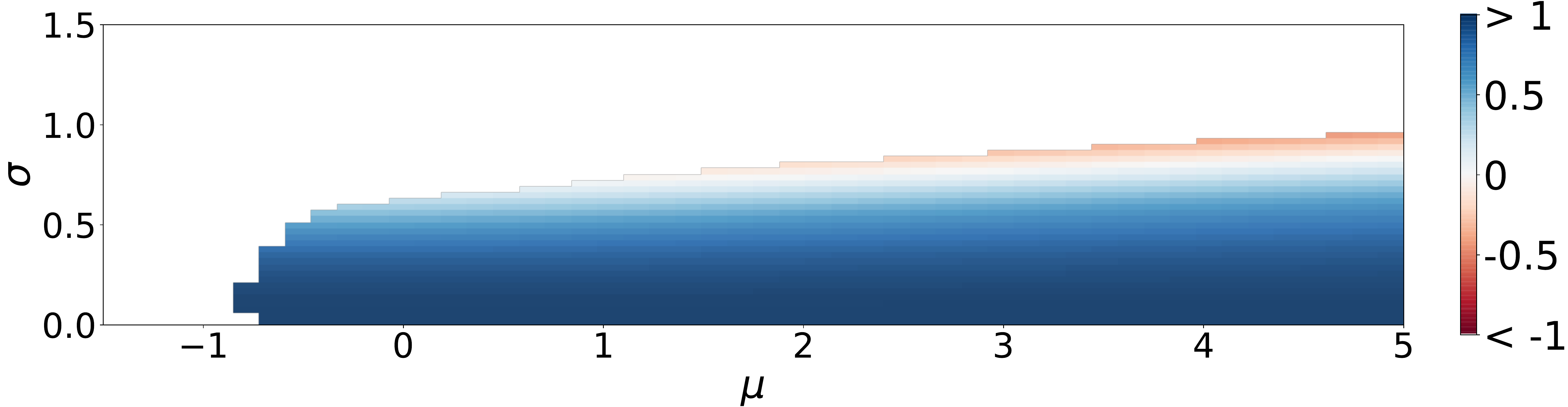}}\\
    \sidesubfloat[]{\includegraphics[width=0.7\textwidth]{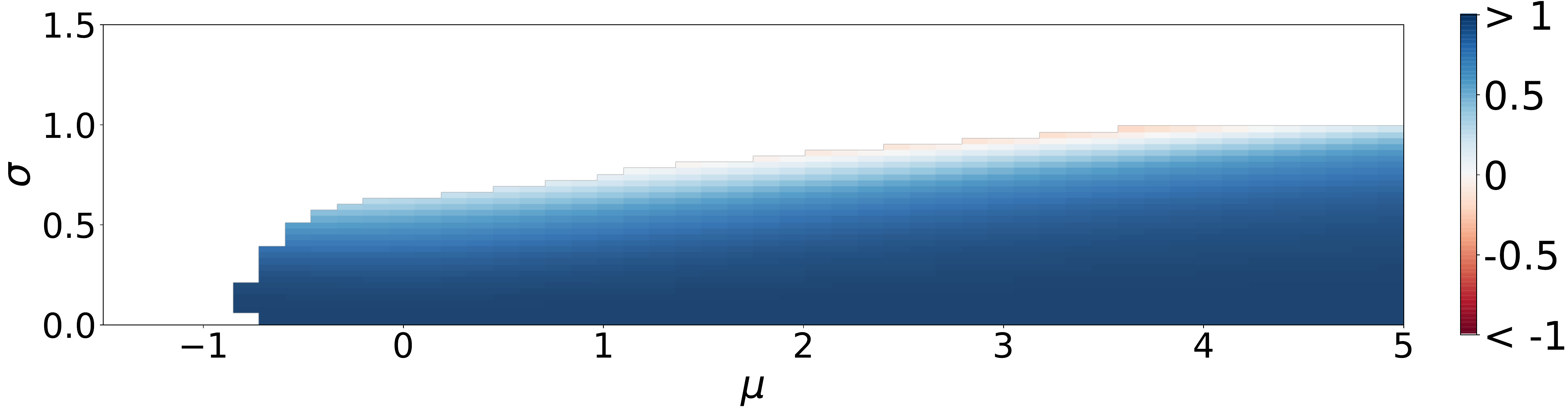}}\\
    \sidesubfloat[]{\includegraphics[width=0.7\textwidth]{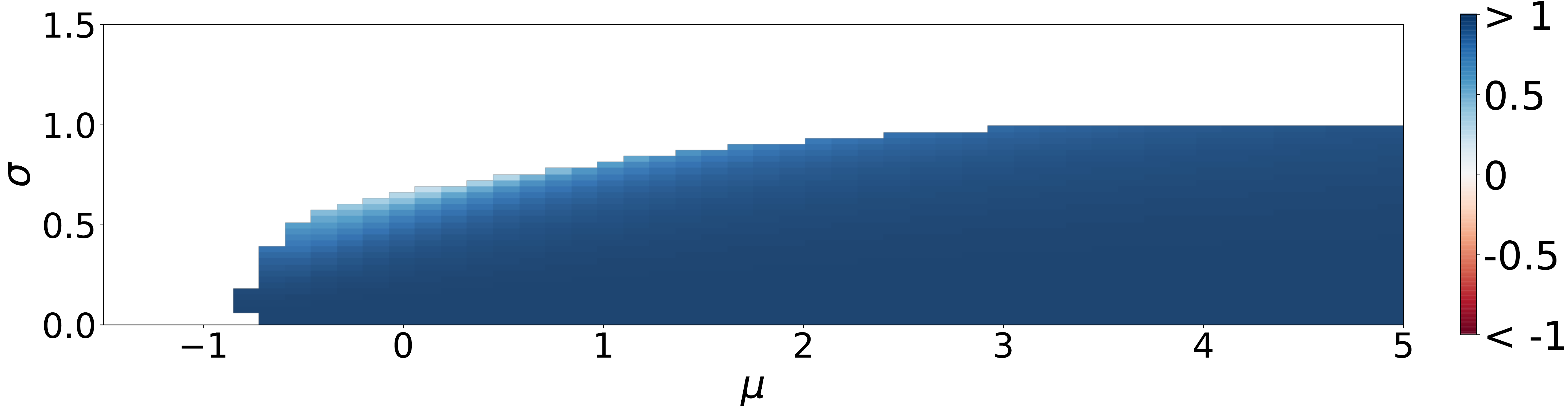}}
    \caption{{\bf RS-RSB phase diagrams over $\mu$-$\sigma$ space.} The right hand side of the replica-symmetry breaking criteria in equation \eqref{stability} is computed over parameter space with $(K,\sigma_K) = (1,0.2)$ and (a) $D = 0$, (b) $D = 1\times 10^{-6}$, (c) $D = 1\times 10^{-3}$, (d) $D = 10^{-1}$, and (e) $D = 1.0$. The blue color corresponds to the stability criteria being satisfied, implying that the replica symmetric solution is stable and the ecosystem exhibits a unique equilibrium. On the other hand, the red region is where the stability criteria is violated, and the ecosystem is in the RSB phase where the ecosystem has multiple equilibrium solutions. Notice that the RSB phase gradually shrinks as one increases the noise level of the system with the region completely disappearing between $D = 0.1$ and $D = 1.0$}
    \label{fig:RSB_phasediagrams}
\end{figure}

Using our numerical cavity solutions to the population moments, we can construct the phase diagram and find the region in parameter space where the ecosystem exhibits multiple equilibria. These diagrams are depicted in Figure \ref{fig:RSB_phasediagrams}.

\floatsetup[figure]{style=plain,subcapbesideposition=top}

\makeatletter\@input{xx.tex}\makeatother